\documentclass[preprint,aps,prd,amsmath,amssymb]{revtex4}
\pdfoutput=1
\usepackage{graphicx}
\usepackage{color}
\usepackage{dcolumn}% Align table columns on decimal point
\usepackage{bm}% bold math
\usepackage{amsmath}
\usepackage{amsfonts}
\usepackage{bbm}
\usepackage{subfigure}
\usepackage{setspace}
\newcommand{\beq}{\begin{eqnarray}}
\newcommand{\eeq}{\end{eqnarray}}

\newcommand{\bmp}{\noindent\begin{minipage}{16cm}}
\newcommand{\emp}{\end{minipage}\vskip 7mm} % 7mm untightened

\def\drawbox#1#2{\hrule height#2pt
        \hbox{\vrule width#2pt height#1pt \kern#1pt
              \vrule width#2pt}
              \hrule height#2pt}

\def\Asym#1#2{\vcenter{\vbox{\drawbox{#1}{#2}
              \kern-#2pt % line up boxes
              \drawbox{#1}{#2}}}}

\begin{document}
%%%%%%%%%%%%%%%%%%%%%%%%%%%%%%%%%%%%%%%%%%%%%%%%%%%%%%%%%%%%%%%%%%%%%%%%%%%
\title{\Large  The Electroweak Phase Transition \\ in\\ 
Nearly Conformal Technicolor}
\author{James M. {\sc Cline}}
\email{jcline@hep.physics.mcgill.ca}
\affiliation{McGill University, Montr\'eal, Qu\'ebec H3A 2T8, Canada.}
\author{Matti  {\sc J\"arvinen}}
\email{mjarvine@ifk.sdu.dk}
\author{Francesco {\sc Sannino}}
\email{sannino@fysik.sdu.dk}
\affiliation{High Energy Center, University of Southern Denmark, Campusvej 55, DK-5230 Odense M, Denmark.}

%%%%%%%%%%%%%%%%%%%%%%%%%%%%%%%%%%%%%%%%%%%%%%%%%%%%%%%%%%%%%%%%%%%%%%%%%%%%%%%%%%%%%%%%%%%%%%%%%%%%%%%%%%%%%%%%%%%%%%%%%%%%%%%%%%%%%%%%%%%%%%

\begin{abstract} We examine the temperature-dependent electroweak
phase transition in extensions of the Standard Model in which the
electroweak symmetry is spontaneously broken via strongly coupled,
nearly-conformal dynamics. In particular, we focus on the low energy
effective theory used to describe Minimal Walking Technicolor at the
phase transition. Using the one-loop effective potential with ring
improvement, we identify significant regions of parameter space which
yield a sufficiently strong first order transition for electroweak
baryogenesis. The composite particle spectrum corresponding to these regions can be produced and studied at the Large Hadron Collider experiment. 
We note the possible emergence of a second phase
 transition at lower temperatures. This occurs when the underlying
 technicolor theory possesses a nontrivial center symmetry.

\end{abstract}

%%%%%%%%%%%%%%%%%%%%%%%%%%%%%%%%%%%%%%%%%%%%%%%%%%%%%%%%%%%%%%%%%%%%%%%%
\maketitle

\section{Introduction}

The experimentally observed baryon asymmetry of the universe may be
generated at the electroweak phase transition (EWPT)
\cite{Shaposhnikov:1986jp,Shaposhnikov:1987tw,Shaposhnikov:1987pf,
Farrar:1993sp,Farrar:1993hn,Gavela:1993ts,Gavela:1994ds}. {}For the
mechanism to be applicable it requires the presence of new physics 
beyond the Standard Model (SM)
\cite{Nelson:1991ab,Joyce:1994bi,Joyce:1994fu,Joyce:1994zn,
Joyce:1994zt,Cline:1995dg}. {An essential condition for electroweak
baryogenesis is that the  baryon-violating interactions induced by
electroweak sphalerons are sufficiently slow immediately after the
phase transition to avoid the destruction of the baryons that have
just been created.  This is achieved when the thermal average of the
Higgs field evaluated on the ground state,  in the broken phase of
the electroweak symmetry, is large enough compared to the critical
temperature at the time of the transition (see for example
ref.~\cite{Cline:2006ts} and references therein),
\beq
        \phi_c/ T_c   > 1.
\label{cond} 
\eeq 
In the SM, the bound (\ref{cond}) was believed to be satisfied
only for very light Higgs
bosons \cite{Carrington:1991hz,Arnold:1992fb,Arnold:1992rz,
Anderson:1991zb,Dine:1992wr}.  However, this was before the mass
of the top quark was known.  With $m_t=175$ GeV, nonperturbative studies
of the phase transition \cite{Kajantie:1995kf} show that 
the bound (\ref{cond}) cannot be satisfied for {\it any} value of the
Higgs mass.
In addition to the
difficulties with producing a large enough initial baryon asymmetry,
the impossibility of satisfying the sphaleron constraint (\ref{cond})
in the SM (Standard Model) provides an incentive for seeing whether the situation
improves in various extensions of the SM. We refer to
\cite{Cline:2006ts} for a summary of the different attempts in this
direction. 

In this paper we explore the electroweak phase transition in a model
in which the electroweak symmetry is broken dynamically
\cite{Weinberg:1979bn,Susskind:1978ms}. A dynamical origin behind the
spontaneous breaking of the electroweak symmetry is a natural
extension of the SM. However, electroweak precision data and
constraints from flavor changing neutral currents both disfavor an
underlying gauge dynamics resembling too closely a scaled-up version
of Quantum Chromodynamics (QCD) (see
\cite{Sannino:2008ha,Hill:2002ap,Lane:2002wv} for recent reviews). 

Since technicolor models have been less fashionable than
supersymmetric models in the last decade, it is worthwhile to review the
recent progress that has enhanced their attractiveness from the
particle physics perspective.  One area of progress  is in the 
understanding of the phase diagram
\cite{Sannino:2004qp,Dietrich:2006cm,Ryttov:2007sr,Ryttov:2007cx}, as
function of the number of flavors and colors, of any SU(N)
non-supersymmetric gauge theory with fermionic matter transforming
according to various representations of the underlying gauge group. 
This has made it possible to provide the first classification of the
possible theories one can use to break the electroweak symmetry
\cite{Dietrich:2005jn,Dietrich:2006cm}.  New analytic tools such as
the all-order beta function \cite{Ryttov:2007cx} allow the
determination, for the first time, of the anomalous dimension of the
mass of the fermions at the nonperturtative infrared fixed point. 
This information is crucial for walking technicolor models
\cite{Holdom:1984sk,Eichten:1979ah,Holdom:1981rm,Yamawaki:1985zg,Appelquist:1986an,Lane:1989ej},
{\it  i.e.}, the ones for which the underlying gauge dynamics is
nearly conformal.

A key realization that enabled further progress was that  gauge
theories with fermions in two-index (symmetric or adjoint)
representations of the underlying gauge group have interesting
features
\cite{Sannino:2004qp,Dietrich:2005jn,Dietrich:2006cm,Ryttov:2007sr,Ryttov:2007cx},
such as the possibility of the existence of a nonperturbative
infrared fixed point for a very low number of flavors
\cite{Sannino:2004qp}, naturally reducing the tension with precision
data \cite{Sannino:2004qp,Dietrich:2005jn,Foadi:2007ue,Foadi:2007se}.
These properties make them intriguing candidates for walking
technicolor type models \cite{Sannino:2004qp,Dietrich:2005jn}  
(related studies can be found in \cite{Christensen:2005cb}).   In
contrast, the naive scaling up of QCD, which is far from conformal,
is strongly contradicted by phenomenological
constraints \footnote{The reader will find in Appendix F of
\cite{Sannino:2008ha} a complete account of alternative large N
limits one can use to gain information on the spectrum of theories
with matter in higher dimensional representation}.  

Another important development occurred in  first principle lattice
simulations of the minimal walking technicolor theories, carried out
in refs.\ \cite{Catterall:2008qk,
Catterall:2007yx,Shamir:2008pb,DelDebbio:2008zf,DelDebbio:2008wb}.
These studies give preliminary support to the analytical arguments
that these theories are near or actually already conformal. The case
of fermions in the fundamental representation has been investigated
in  \cite{Catterall:2008qk,Appelquist:2007hu,Deuzeman:2008sc}.

On the astrophysical side, technicolor models are capable of
providing interesting dark matter candidates, since the  new strong 
interactions confine techniquarks in technimeson and technibaryon
bound states. The spin of the technibaryons depends on the
representation according to which the technifermions transform, and
the numbers of flavors and colors. The lightest technimeson is
short-lived, thus evading BBN constraints, but the lightest
technibaryon has typically \footnote{there may be situations in which
the technibaryon is a goldstone boson of an enhanced flavor
symmetry} a mass of the order 
\begin{equation}
 m_{TB}  \sim  1-2\ {\rm TeV} \ .
\end{equation} 

Technibaryons are therefore natural dark matter candidates 
\cite{Nussinov:1985xr,Barr:1990ca,Gudnason:2006yj}. In fact it is
possible to {naturally} understand the observed ratio of the dark to
luminous matter mass fraction of the universe if the technibaryon
possesses an asymmetry 
\cite{Nussinov:1985xr,Barr:1990ca,Gudnason:2006yj}. If the latter is
due to a net $B-L$ generated at some high energy scale, then this
would be subsequently distributed among {\em all} electroweak
doublets by fermion-number violating processes in the SM at
temperatures above the electroweak scale
\cite{Shaposhnikov:1991cu,Kuzmin:1991ft,Shaposhnikov:1991wi}, thus
naturally generating a technibaryon asymmetry as well. To avoid
experimental constraints the technibaryon should be constructed in
such a way as to be a complete singlet under the electroweak
interactions \cite{Barr:1990ca,Dietrich:2006cm} while still having a
nearly conformal underlying gauge theory \cite{Dietrich:2006cm}. In
this case it would be hard to detect it in current earth-based
experiments  such as CDMS
\cite{Bagnasco:1993st,Gudnason:2006yj,Kouvaris:2008hc,Akerib:2004fq,Akerib:2005kh}. 
Other possibilities have been envisioned in
\cite{Kouvaris:2007iq,Khlopov:2007ic} and possible astrophysical
effects studied in \cite{Kouvaris:2007ay}.  One can  alternatively
obtain dark matter from possible associated new sectors instead of
the technicolor sector \cite{Kainulainen:2006wq}, including those
which are not gauged under the electroweak interactions
\cite{Dietrich:2006cm}.   In \cite{Sannino:2008ha} the reader will
find an up-to-date summary of the recent efforts in this direction.

Coming to the main topic of this paper,  the order of the electroweak
phase transition (EWPT) depends on the underlying type of strong dynamics
and  plays an important role for baryogenesis
\cite{Cline:2002aa,Cline:2006ts}. The technicolor chiral phase
transition at finite temperature is mapped onto the electroweak one.
Attention must be paid to the way in which the electroweak symmetry
is embedded into the global symmetries of the underlying technicolor
theory.  An interesting preliminary analysis dedicated to earlier
models of technicolor has been performed in \cite{Kikukawa:2007zk}. 

In this work, we wish to investigate the EWPT in a class of realistic
and viable technicolor models. An explicit phenomenological
realization of walking models consistent with the electroweak
precision data   is termed Minimal Walking Technicolor
(MWT) \cite{Foadi:2007ue}. It is based on an SU(2) gauge theory
coupled to two flavors of adjoint techniquarks. This model is thought
to lie close, in theory space, to theories with nontrivial infrared
fixed points \cite{Sannino:2004qp,Ryttov:2007cx}. Indeed it is
possible that it already has such a fixed point itself. In the
vicinity of such a zero of the beta function, the coupling constant
flows slowly (``walks''). This theory possesses an SU(4) global
symmetry. At the LHC one will observe the composite states which are
classified according to irreducible representations of the stability
group left invariant by the technifermion condensate. The stability
group, here, corresponds to the SO(4) symmetry which contains the
SU(2)  custodial symmetry of the SM. We choose the natural SM
embedding, as detailed in the following section.

In ref.\ \cite{Foadi:2007ue} a comprehensive Lagrangian was
introduced for this model, taking into account the global symmetries
of the underlying gauge theory, the walking dynamics via the modified
Weinberg sum rules \cite{Appelquist:1998xf}, and the constraints
coming from precision data \cite{Foadi:2007se}. The effective theory
contains composite scalars and spin-one vectors. Compatibility
between the electroweak precision constraints and tree-level
unitarity of $WW$-scattering was demonstrated  in
\cite{Foadi:2008ci}.

 The study of
longitudinal WW scattering unitarity versus precision measurements
within the effective Lagrangian approach demonstrated that it is
possible to pass the precision tests while simultaneously delaying
the onset of unitarity \cite{Foadi:2008ci}. 

In the present work we will use as a template the low energy
effective theory developed in  \cite{Foadi:2007ue}. We start in
section \ref{sect2} by summarizing the basic theory, highlighting the
degrees of freedom relevant near the phase transition.  In section
\ref{sect3} the finite-temperature effective potential is then
computed at the one-loop order, including the resummation of ring
diagrams.  Our analysis is presented in section \ref{sect4}.   As a
preliminary investigation we adopt the high-temperature expansion
results for the effective potential.  We then explore the region of
the effective theory parameters yielding a first-order phase
transition and study its strength. The ratio of the composite Higgs
thermal expectation value at the critical temperature divided by the 
corresponding temperature is determined as function of the parameters
of the low energy effective theory. We identify a significant region
of parameter space where this ratio is sufficiently large to induce
electroweak baryogenesis. {The spectrum of the composite spin-zero states directly associated to these regions can be investigated and the related particles produced at the Large Hadron Collider experiment. In the subsection \ref{sect4.d} we note  the possible emergence of a second phase
 transition at lower temperatures, {\it i.e.}, the confinement/deconfinement one. This transition occurs when the underlying
 technicolor theory possesses a nontrivial center symmetry.} Several appendices are provided, which give
details concerning our analytical results.

\section{Introducing Minimal Walking Technicolor}
\label{sect2}

\subsection{The underlying degrees of freedom and Lagrangian}

The new dynamical sector we consider, which underlies the Higgs
mechanism, is an SU(2) technicolor gauge theory with two adjoint
technifermions \cite{Sannino:2004qp}. The two adjoint fermions may be
written as \beq Q_L^a=\left(\begin{array}{c} U^{a} \\D^{a}
\end{array}\right)_L , \qquad U_R^a \ , \quad D_R^a \ ,  \qquad
a=1,2,3 \ ,\eeq with $a$ being the adjoint color index of SU(2). The
left-handed fields are arranged in three doublets of the SU(2)$_L$
weak interactions in the standard fashion. The condensate is $\langle
\bar{U}U + \bar{D}D \rangle$ which correctly breaks the electroweak
symmetry. The model as described so far suffers from the Witten
topological anomaly \cite{Witten:1982fp}. However, this can easily be
addressed by adding a new weakly charged fermionic doublet which is a
technicolor singlet \cite{Dietrich:2005jn}. Schematically,
\beq L_L =
\left(
\begin{array}{c} N \\ E \end{array} \right)_L , \qquad N_R \ ,~E_R \
. \eeq In general, the gauge anomalies cancel using the 
generic hypercharge assignment
\begin{align}
Y(Q_L)=&\frac{y}{2} \ ,&\qquad Y(U_R,D_R)&=\left(\frac{y+1}{2},\frac{y-1}{2}\right) \ , \label{assign1} \\
Y(L_L)=& -3\frac{y}{2} \ ,&\qquad
Y(N_R,E_R)&=\left(\frac{-3y+1}{2},\frac{-3y-1}{2}\right) \ \label{assign2} ,
\end{align}
where the parameter $y$ can take any real value \cite{Dietrich:2005jn}. In our notation
the electric charge is $Q=T_3 + Y$, where $T_3$ is the weak
isospin generator. One recovers the SM hypercharge
assignment for $y=1/3$.

To discuss the symmetry properties of the
theory it is
convenient to use the Weyl basis for the fermions and arrange them 
in a vector transforming according to the
fundamental representation of SU(4),
\beq Q= \begin{pmatrix}
U_L \\
D_L \\
-i\sigma^2 U_R^* \\
-i\sigma^2 D_R^*
\end{pmatrix},
\label{SU(4)multiplet} \eeq where $U_L$ and $D_L$ are the left-handed
 techniup and technidown, respectively, and $U_R$ and $D_R$ are
the corresponding right-handed particles. Assuming the standard
breaking to the maximal diagonal subgroup, the SU(4) symmetry
spontaneously breaks to SO(4). This is driven by the
condensate 
\beq \langle Q_i^\alpha Q_j^\beta
\epsilon_{\alpha \beta} E^{ij} \rangle =-2\langle \overline{U}_R U_L
+ \overline{D}_R D_L\rangle \ , \label{conde}
 \eeq
where the indices $i,j=1,\ldots,4$ denote the components
of the tetraplet of $Q$, and the Greek indices indicate the ordinary
spin. The matrix $4\times 4$ $E$ is defined in terms
of the 2-dimensional unit matrix by
 \beq E=\left(
\begin{array}{cc}
0 & \mathbbm{1} \\
\mathbbm{1} & 0
\end{array}
\right) \ , \eeq
the antisymmetric tensor is
$\epsilon_{\alpha \beta}=-i\sigma_{\alpha\beta}^2$ and we used $\langle
 U_L^{\alpha} {{U_R}^{\ast}}^{\beta} \epsilon_{\alpha\beta} \rangle=
 -\langle  \overline{U}_R U_L
 \rangle$. A similar expression holds for the $D$ techniquark.
The above condensate is invariant under an SO(4) symmetry. 
This yields nine broken  generators with associated Goldstone bosons.

Replacing the Higgs sector of the SM with MWT, one writes
\begin{eqnarray}
\mathcal{L}_H &\rightarrow &  -\frac{1}{4}{\cal F}_{\mu\nu}^a {\cal F}^{a\mu\nu} + i\bar{Q}_L
\gamma^{\mu}D_{\mu}Q_L + i\bar{U}_R \gamma^{\mu}D_{\mu}U_R +
i\bar{D}_R \gamma^{\mu}D_{\mu}D_R \nonumber \\
&& +i \bar{L}_L \gamma^{\mu} D_{\mu} {L}_L +
i\bar{N}_R \gamma^{\mu}D_{\mu}N_R + i\bar{E}_R
\gamma^{\mu}D_{\mu}E_R
\end{eqnarray}
with the technicolor field strength ${\cal F}_{\mu\nu}^a =
\partial_{\mu}{\cal A}_{\nu}^a - \partial_{\nu}{\cal A}_{\mu}^a + g_{TC} \epsilon^{abc} {\cal A}_{\mu}^b
{\cal A}_{\nu}^c,\ a,b,c=1,\ldots,3$.
For the left-handed techniquarks the covariant derivative is
\begin{eqnarray}
D_{\mu} Q^a_L &=& \left(\delta^{ac}\partial_{\mu} + g_{TC}{\cal
A}_{\mu}^b \epsilon^{abc} - i\frac{g}{2} \vec{W}_{\mu}\cdot
\vec{\tau}\delta^{ac} -i g'\frac{y}{2} B_{\mu} \delta^{ac}\right)
Q_L^c \ .
\end{eqnarray}
Here ${\cal A}_{\mu}$ are the techni gauge bosons, $W_{\mu}$ are the
gauge bosons associated to SU(2)$_L$ and $B_{\mu}$ is the gauge
boson associated to the hypercharge. $\tau^a$ are the Pauli matrices
and $\epsilon^{abc}$ is the fully antisymmetric symbol. In the case
of right-handed techniquarks the third term containing the weak
interactions disappears and the hypercharge $y/2$ has to be replaced
according to whether it is an up or down techniquark. For the
left-handed leptons the second term containing the technicolor
interactions disappears and $y/2$ changes to $-3y/2$. Only the last
term is present for the right-handed leptons with an appropriate
hypercharge assignment.

\subsection{Tree Level Low Energy Theory for MWT}

In \cite{Foadi:2007ue} we constructed the effective theory for MWT
including composite scalars and vector bosons, their
self-interactions, and their interactions with the electroweak gauge
fields and the SM fermions. We have also used the Weinberg  modified
sum rules to constrain the low
energy effective theory. This extension of the SM was thereby shown 
to pass the electroweak precision tests. Near the finite
temperature phase transition the relevant degrees of freedom are the
scalars and hence we will not consider the vector spectrum nor that of
the composite fermions.

\subsubsection{Scalar Sector} \label{sec:scalar}

The relevant effective theory for the Higgs sector at the electroweak
scale consists, in our model, of a composite Higgs and its
pseudoscalar partner, as well as nine pseudoscalar Goldstone bosons
and their scalar partners. These can be assembled in the matrix
\begin{eqnarray}
M = \left[\frac{\sigma+i{\Theta}}{2} + \sqrt{2}(i\Pi^a+\widetilde{\Pi}^a)\,X^a\right]E \ ,
\label{M}
\end{eqnarray}
which transforms under the full SU(4) group according to
\begin{eqnarray}
M\rightarrow uMu^T \ , \qquad {\rm with} \qquad u\in {\rm SU(4)} \ .
\end{eqnarray}
The $X^a$'s, $a=1,\ldots,9$ are the generators of the SU(4) 
group which do not leave  the vacuum expectation value (VEV) of $M$ 
invariant.  $\langle M \rangle$ is given by
\begin{eqnarray}
\langle M \rangle = \frac{v}{2}E
 \ .
\end{eqnarray}
We note that $\sigma$ is a scalar
while the $\Pi^a$'s are pseudoscalars. It is convenient to
separate the fifteen generators of SU(4) into the six that leave the
vacuum invariant, $S^a$, and the remaining nine that do not, $X^a$.

The connection between the composite scalars and the underlying
techniquarks can be derived from their transformation properties under
SU(4), by observing that the elements of the matrix $M$ transform
like techniquark bilinears,
\begin{eqnarray}
M_{ij} \sim Q_i^\alpha Q_j^\beta \varepsilon_{\alpha\beta} \quad\quad\quad {\rm with}\ i,j=1\dots 4.
\label{M-composite}
\end{eqnarray}

The electroweak subgroup can be embedded in SU(4), as explained in detail in \cite{Appelquist:1999dq}. The generators $S^a$, with $a=1,2,3$, form a vectorial SU(2) subgroup of SU(4), which is denoted by SU(2)$_{\rm V}$, while $S^4$ forms a U(1)$_{\rm V}$ subgroup. The $S^a$ generators, with $a=1,..,4$, together with the $X^a$ generators, with $a=1,2,3$, generate an SU(2)$_{\rm L}\times$SU(2)$_{\rm R}\times$U(1)$_{\rm V}$ algebra. This is seen by changing genarator basis from $(S^a,X^a)$ to $(L^a,R^a)$, where
\begin{eqnarray}
L^a \equiv \frac{S^a + X^a}{\sqrt{2}} = \begin{pmatrix}\frac{\tau^a}{2}\ \ \  & 0 \\ 0 & 0\end{pmatrix} \ , \ \
{-R^a}^T \equiv \frac{S^a-X^a}{\sqrt{2}}  = \begin{pmatrix}0 & 0 \\ 0 & -\frac{{\tau^a}^T}{2}\end{pmatrix} \ ,
\end{eqnarray}
with $a=1,2,3$. The electroweak gauge group is then obtained by 
gauging ${\rm SU(2)}_{\rm L}$ and the ${\rm U(1)}_Y$ subgroup of ${\rm SU(2)}_{\rm R}\times {\rm U(1)}_{\rm V}$, where
\begin{eqnarray}
Y =  -{R^3}^T + \sqrt{2}\ Y_{\rm V}\ S^4 \ ,
\end{eqnarray}
and $Y_{\rm V}$ is the U(1)$_{\rm V}$ charge. For example, from 
eqs.~(\ref{assign1}) and (\ref{assign2}) we see that $Y_{\rm V}=y$ for the techniquarks, and $Y_{\rm V}=-3y$ for the new leptons. As SU(4) spontaneously breaks to SO(4), ${\rm SU(2)}_{\rm L}\times {\rm SU(2)}_{\rm R}$ breaks to ${\rm SU(2)}_{\rm V}$. As a consequence, the electroweak symmetry breaks to ${\rm U(1)}_Q$, where
\begin{eqnarray}
Q = \sqrt{2}\ S^3 + \sqrt{2}\ Y_{\rm V} \ S^4 \ .
\end{eqnarray}
The ${\rm SU(2)}_{\rm V}$ group, being entirely contained 
in the unbroken SO(4), acts as a custodial isospin, 
which insures that the $\rho$ parameter is equal to one at tree level.

The electroweak covariant derivative for the $M$ matrix is
\begin{eqnarray}
D_{\mu}M =\partial_{\mu}M - i\,g \left[G_{\mu}(y)M + MG_{\mu}^T(y)\right]  \
, \label{covariantderivative}
\end{eqnarray}
where
\begin{eqnarray}
g\ G_{\mu}(Y_{\rm V}) & = & g\ W^a_\mu \ L^a + g^{\prime}\ B_\mu \ Y  \nonumber \\
& = & g\ W^a_\mu \ L^a + g^{\prime}\ B_\mu \left(-{R^3}^T+\sqrt{2}\ Y_{\rm V}\ S^4\right) \ .
\label{gaugefields}
\end{eqnarray}
Notice that in the last equation, $G_\mu(Y_{\rm V})$ is written for a
general U(1)$_{\rm V}$ charge $Y_{\rm V}$, while in
eq.~(\ref{covariantderivative}) we have to take the U(1)$_{\rm V}$
charge of the techniquarks, $Y_{\rm V}=y$, since these are the
constituents of the matrix $M$, as explicitly shown in
eq.~(\ref{M-composite}).

Three of the nine Goldstone bosons associated with the broken
generators become the longitudinal degrees of freedom of the massive
weak gauge bosons, while the extra six Goldstone bosons will acquire
a mass due to extended technicolor interactions (ETC) as well as the
electroweak interactions {\it per se}. Using a bottom-up approach, we
will not commit to a specific ETC theory, but rather
limit ourselves to
introducing the minimal low energy operators  needed to construct a
phenomenologically viable theory. The new Higgs Lagrangian is
\begin{eqnarray}
{\cal L}_{\rm Higgs} &=& \frac{1}{2}{\rm Tr}\left[D_{\mu}M D^{\mu}M^{\dagger}\right] - {\cal V}(M) + {\cal L}_{\rm ETC} \ ,
\end{eqnarray}
where the potential reads
\begin{eqnarray} \label{Vdef}
{\cal V}(M) & = & - \frac{m^2}{2}{\rm Tr}[MM^{\dagger}] +\frac{\lambda}{4} {\rm Tr}\left[MM^{\dagger} \right]^2 
+ \lambda^\prime {\rm Tr}\left[M M^{\dagger} M M^{\dagger}\right] \nonumber \\
& - & 2\lambda^{\prime\prime} \left[{\rm Det}(M) + {\rm Det}(M^\dagger)\right] \ ,
\end{eqnarray}
and ${\cal L}_{\rm ETC}$ contains all terms which are generated by the ETC interactions, and not by the chiral symmetry breaking sector. 

We explicitly break the SU(4) symmetry in order to provide mass to the Goldstone bosons which are not eaten by the weak gauge bosons. 
Assuming parity invariance,
\begin{eqnarray} \label{VETCdef}
{\cal L}_{\rm ETC} = \frac{m_{\rm ETC}^2}{4}\ {\rm Tr}\left[M B M^\dagger B + M M^\dagger \right] + \cdots \ ,
\end{eqnarray}
where the ellipses represent possible higher dimensional operators, and $B\equiv 2\sqrt{2}S^4$ commutes with the SU(2)$_{\rm L}\times$SU(2)$_{\rm R}\times$U(1)$_{\rm V}$ generators.

The potential ${\cal V}(M)$ is SU(4) invariant. It produces a VEV
which parameterizes the techniquark condensate, and spontaneously
breaks SU(4) to SO(4). In terms of the model parameters the VEV is
\begin{eqnarray}
v^2=\langle \sigma \rangle^2 = \frac{m^2}{\lambda + \lambda^\prime - \lambda^{\prime\prime} } \ ,
\label{VEV}
\end{eqnarray}
while the Higgs mass is
\begin{eqnarray}
M_H^2 = 2\ m^2 \ .
\end{eqnarray}
The linear combination $\lambda + \lambda^{\prime} -
\lambda^{\prime\prime}$ corresponds to the Higgs self-coupling in
the SM. The three pseudoscalar mesons $\Pi^\pm$, $\Pi^0$, correspond
to the three massless Goldstone bosons which are absorbed by the
longitudinal degrees of freedom of the $W^\pm$ and $Z$ boson. The
remaining six uneaten Goldstone bosons are technibaryons, and all
acquire tree-level degenerate masses through (not yet specified) ETC interactions:
\begin{eqnarray}
M_{\Pi_{UU}}^2 = M_{\Pi_{UD}}^2 = M_{\Pi_{DD}}^2 = m_{\rm ETC}^2  \ .
\end{eqnarray}
The remaining scalar and pseudoscalar masses are
\begin{eqnarray}
M_{\Theta}^2 & = & 4 v^2 \lambda^{\prime\prime} \nonumber \\
M_{A^\pm}^2 = M_{A^0}^2 & = & 2 v^2 \left(\lambda^{\prime}+\lambda^{\prime\prime}\right)
\end{eqnarray}
for the technimesons, and
\begin{eqnarray}
M_{\widetilde{\Pi}_{UU}}^2 = M_{\widetilde{\Pi}_{UD}}^2 = M_{\widetilde{\Pi}_{DD}}^2 =
m_{\rm ETC}^2 + 2 v^2 \left(\lambda^{\prime} + \lambda^{\prime\prime }\right) \ ,
\end{eqnarray}
for the technibaryons. 
Ref.\ \cite{Hong:2004td} provides further insight into some of these
mass relations.

\subsubsection{Fourth Lepton Family and Yukawa Interactions}

The fermionic content of the effective theory consists of the SM
quarks and leptons, the new lepton doublet $L=(N,E)$ introduced to
cure the Witten anomaly, and a composite techniquark-technigluon
doublet. In fact the most relevant contributions are the  ones of the
top quark and the new lepton contribution due to their large Yukawa
couplings and their relatively small zero-temperature masses,
compared to the EWPT temperature.

Many  extensions of technicolor have been suggested in the literature
to  provide masses to ordinary fermions.  Some of the extensions use
additional strongly coupled gauge dynamics, while others introduce
fundamental scalars. Many variants of the schemes presented above
exist, and a review of the major models is given by Hill and Simmons
\cite{Hill:2002ap}. At the moment there is not yet a consensus on
which ETC is the best. To keep the number of fields minimal,  ref.\
\cite{Foadi:2007ue} made the most economical ansatz, {\it i.e.},
ignorance of the complete ETC theory was parametrized by simply
coupling the fermions to the low energy effective composite Higgs.
This simple construction minimizes the flavor-changing neutral
current problem. It is worth mentioning that it is possible to
engineer a schematic ETC model proposed first by Randall in
\cite{Randall:1992vt} and adapted for the MWT in \cite{Evans:2005pu}
for which the effective theory presented here can be
considered as a minimal description \footnote{Another nonminimal way
to give masses to the ordinary fermions is to (re)introduce a new
Higgs doublet as already done many times in the literature
\cite{Simmons:1988fu,Dine:1990jd,Kagan:1990az,Carone:1992rh,Carone:1993xc,Gudnason:2006mk}.
This possibility and its phenomenological applications will be
studied elsewhere.}. The details can be found in \cite{Foadi:2007ue}.
In our study of the phase transition  we will not consider the
composite fermions since they are expected to be much heavier than
the scalar degrees of freedom.

\section{MWT  - Effective Potential}
\label{sect3}

The tree-level effective potential is obtained by evaluating the
potential in (\ref{Vdef}) and (\ref{VETCdef}) in the background where
the Higgs fields assumes the vacuum expectation value $\sigma$, 
{\it i.e.,}
$M = \sigma E/2$. It has the SM form
\begin{eqnarray} \label{Vtree}
 V^{(0)} = \frac{1}{4}\left(\lambda+\lambda'-\lambda''\right) \left(\sigma^2-v^2\right)^2 =\frac{M_H^2}{8v^2} \left(\sigma^2-v^2\right)^2 \ .
\end{eqnarray}
The effective potential at one loop can be naturally divided into
zero- and nonzero-temperature contributions.

\subsection{Zero Temperature Contribution} \label{zeroTV}

We begin by constructing the one-loop effective potential at zero
temperature.  We fix the counterterms so as to  preserve the
tree-level definitions of the VEV and the Higgs mass, {\it i.e.,}
$M^2_H = 2\bar{\lambda} v^2$ with $\bar{\lambda}=\lambda +
\lambda^{\prime} -\lambda^{\prime \prime}$. The one loop contribution
to the potential then reads:
\begin{eqnarray} \label{VT0}
V^{(1)}_{T=0} = \frac{1}{64\pi^2} \sum_{i} n_i\, f_{i}(M_i(\sigma)) +
V_{\rm GB} \ ,
\end{eqnarray}
where the index $i$ runs over all of the mass eigenstates, 
except for the Goldstone bosons (GB), and $n_i$ is the multiplicity 
factor for a given scalar particle while for Dirac fermions is $-4$ 
times the multiplicity factor of the specific fermion.  
The function $f_i$ is:
\begin{equation} \label{fdef}
f_i = M^4_i(\sigma) \left[\log\frac{M^2_{i}(\sigma)}{M^2_i(v)}  - \frac{3}{2}\right] + 2M^2_i(\sigma) \, M^2_i(v)  \ ,
\end{equation}
where $M^2_i(\sigma)$ is the background dependent mass term of the
$i$-th particle. This prescription would lead to  infrared
divergences in the 't Hooft-Landau gauge for $V_{\rm GB}$, the GB
contribution, when evaluated at the tree-level VEV, due to the
vanishing of the GB masses. Different ways of dealing with this
problem have been discussed in the literature. One possibility is to
regularize the infrared divergence by replacing $M^2_i(v)$ with some
characteristic mass scale.   However with this prescription the
tree-level VEV and Higgs mass get shifted by the presence of the
one-loop correction.  A simpler approach is to neglect the GB
contribution, since in practice it never has a strong effect on the
phase transition.  We tried both methods and found that they give
essentially indistinguishable results.

To explicitly evaluate the potential above it is useful to split the
scalar matrix into four $2\times 2$ blocks as follows:
\begin{equation}  
M=\begin{pmatrix} {\cal X}  & {\cal O}  \\ {\cal O}^T & {\cal Z} \end{pmatrix} \ ,
\end{equation}
with ${\cal X}$ and ${\cal Z}$ two complex symmetric matrices
accounting for six independent degrees of freedom each and ${\cal O}$
a generic complex $2\times 2$ matrix representing eight real bosonic
fields. ${\cal O}$ accounts for the SM-like Higgs doublet, a second
doublet, and the three GB's absorbed by the
longitudinal  gauge bosons. We find $n_{\cal X} = n_{\cal Z} = 6$
while the two weak doublets split into two SU(2)$_V$ isoscalars, {\it i.e.,}
the Higgs ($n_H=1$) and ${\Theta}$ ($n_{\Theta}=1$) with different
masses and two independent triplets, {\it i.e.,} $n_{GB}=3$ and $n_{A}=3$. 
In Appendix~\ref{AppA} we summarize the tree-level expressions for
the background-dependent masses of the scalar states. 

For the contribution of the gauge bosons we have $n_W=6$ and $n_Z=3$.
In the fermionic sector we will consider only the heaviest
particles, {\it i.e.,} the top for which $n_T = -12$ and the two new
leptons $n_N = n_E = -4$.  

\subsection{One Loop Finite Temperature Effective Potential} \label{sec:oneloopT}
The one-loop, ring-improved, finite-temperature effective potential 
can be divided into fermionic, scalar and vector contributions,
\begin{eqnarray}
V_T^{(1)} = {V_T^{(1)}}_{\rm f}+ {V_T^{(1)}}_{\rm b}+ {V_T^{(1)}}_{\rm gauge}\ .
\end{eqnarray}
The fermionic contribution at high temperature reads:
\begin{eqnarray} \label{VTf}
{V_T^{(1)}}_{\rm f} = 2\frac{T^2}{24} \sum_{f} n_{f} M_{f}^2(\sigma) +\frac{1}{16 \pi^2}\sum_fn_{f} M_{f}^4(\sigma)\left[\log\frac{M_{f}^2(\sigma)}{T^2}-c_f\right]
\end{eqnarray}
where $c_f \simeq 2.63505$, $n_{\rm Top}=3$, $n_{N}=n_{E}=1$, and we 
have neglected ${\cal O}\left(1/T^2\right)$ terms.
The field-dependent masses are
\begin{equation}
M_{\rm Top} (\sigma) = m_{\rm Top}\frac{\sigma}{v} \ , \quad  M_{\rm N}(\sigma) = m_{ N}\frac{\sigma}{v} \ , \quad M_{\rm E} = m_{E}\frac{\sigma}{v} \ ,
\end{equation}
with $m_{\rm Top}$, $m_{N}$ and $m_{E}$ the physical masses. 
Notice that the logarithmic term in (\ref{VTf}) combines with a 
similar term in the zero-temperature potential (\ref{VT0}) so that their sum is analytic in the masses $M_{f}^2(\sigma)$.
 
For the scalar part of the thermal potential one must resum the
contribution of the ring diagrams. Following Arnold and Espinosa
\cite{Arnold:1992rz} we write
\begin{eqnarray} \label{VTb}
{V_T^{(1)}}_{\rm b} =\frac{T^2}{24} \sum_{b} n_{b} M_{b}^2(\sigma)  - \frac{T}{12\pi} \sum_b\,n_b \,M_{b}^3(\sigma,T) \nonumber\\ 
-\frac{1}{64 \pi^2}\sum_bn_{b} M_{b}^4(\sigma)\left[\log\frac{M_{b}^2(\sigma)}{T^2}-c_b\right]
\ ,
\end{eqnarray}
where $c_b\simeq 5.40762 $ and $M_b(\sigma,T)$ the thermal mass
which follows from the 
tree-level plus one-loop thermal contribution to the potential
(see Appendix~\ref{AppA}). {}For the gauge bosons, 
\begin{eqnarray} \label{VTgb}
{V_T^{(1)}}_{\rm gauge} =\frac{T^2}{24} \sum_{gb} 3 M_{gb}^2(\sigma)  - \frac{T}{12\pi} \sum_{gb}\left[ 2 M_{T,gb}^3(\sigma)+ M_{L,gb}^3(\sigma,T)\right] \nonumber \\
-\frac{1}{64 \pi^2}\sum_{gb}n_{gb} M_{gb}^4(\sigma)\left[\log\frac{M_{gb}^2(\sigma)}{T^2}-c_b\right] \ .
\end{eqnarray}
Here $M_{T,gb}$ ($M_{L,gb}$ ) is the transverse (longitudinal) mass
of a given gauge boson and we have $M_{T,gb}(\sigma)=
M_{L,gb}(\sigma,T=0) = M_{gb}(\sigma)$. Only the longitudinal gauge
bosons acquire a thermal mass squared at the leading order, 
$O(g^2T^2)$. The transverse bosons acquire instead a 
magnetic mass squared  of order $g^4T^2$ which we have neglected. 

The explicit form of the transverse and longitudinal gauge boson 
mass matrix is given in Appendix~\ref{AppB}. 

\section{Results}  
\label{sect4}

We used the one-loop high temperature approximation together with the
summation of the ring-diagrams  to evaluate the effective potential
in our numerical calculations. The full expression of the finite
temperature potential is given as a sum of the tree level potential
(\ref{Vtree}), the zero-temperature one-loop contribution
(\ref{VT0}), and the one-loop thermal corrections at high
temperature, (\ref{VTf}), (\ref{VTb}), and (\ref{VTgb}). We assumed
that the phase transition takes place when the two minima are
degenerate. This then defines the critical value of the thermal
average of the composite Higgs field $\phi_c$, in the broken phase,
at the critical temperature $T_c$. Above the critical temperature the
ground state is the one at the origin of the Higgs field. {}For
convenience we subtracted from the potential a temperature-dependent
constant which is defined in such a way that $V(\sigma,T)=0$ for
$\sigma=0$.

The relevant input parameters are the zero-temperature masses of the
Higgs ($M_H$) and its pseudoscalar partner $\Theta$ ($M_\Theta$). The
phase transition also depends on the masses of the scalar partners of
the Goldstone bosons $A^{0,\pm}$ ($M_A$), on the mass scale of the
scalar baryons $m_{\rm ETC}$, and on the masses of the heavy
fermions. For simplicity, we choose the masses of the new fermions to
be equal, \begin{eqnarray} M_{E}^2 = M_{N}^2 \equiv M_{\rm f}^2 \ .
\end{eqnarray} This choice does not seem to have a strong effect on
the phase transition; for example we checked that using instead $M_E
\simeq 2 M_N$, very similar results were obtained. We have neglected
the heavy composite vectors of MWT since they are expected to
decouple at the scale of the EWPT.   At this scale, the couplings to
the SM gauge bosons are simply $g$, $g'$. We set the parameter $y$ to
$y=1/3$ so that the MWT hypercharge assignment equals the SM one.
Notice that $y$ appears only in the longitudinal Debye mass of the
$Z$ boson. Since the effective potential terms are proportional to
$M_i^{2}(\sigma)$ or $M_i^{4}(\sigma)$, the contributions of the
fermions and the composite scalars typically dominate over that of
the relatively light $Z$ boson, whence the dependence of the phase
transition on $y$ is negligible.

It is instructive to consider two limiting cases, for which the
thermal mass spectrum simplifies: light and heavy ETC masses.  
Interpolating between these two cases would require some way of
smoothly connecting the thermal masses when the heavy ETC states have
decoupled, to those for which they are fully contributing. We
discuss these separately in the following subsections.

\subsection{Heavy ETC-induced masses}

We first consider $\phi_c/T_c$ in the heavy ETC mass scenario, {\it
i.e.,} taking the limit $m_{\rm ETC}/T_c \gg 1$. When the scalar
baryons become heavy their contributions to the effective potential
become negligible. Since $\phi_c/T_c$ is more sensitive to $M_H$ and
$M_\Theta$ than to the other masses, we chose to plot it  in the
($M_H,M_{\Theta}$) plane, while varying the remaining parameters
$M_A$ and $M_{\rm f}$.  The resulting dependences are shown in 
in fig.~\ref{fig:heavymetc}. The contour values
of $\phi_c/T_c$  are $\phi_c/T_c=0.5,\ldots, 3.0$ from lighter to
darker shades with steps of $0.5$. Recall that electroweak
baryogenesis requires $\phi_c/T_c \gtrsim 1$.  

In the triangular regions in the upper left corners of the plots, the
broken phase is metastable already at $T=0$, whence there is no phase
transition. When one approaches this region from below one observes
that $T_c$ goes to zero and $\phi_c/T_c$ blows up. This happens since
the one-loop zero-temperature potential induces an almost degenerate
minimum at the origin together with the one at a finite value of
$\phi$. At this point any small temperature favors the minimum at the
origin. Such small temperatures are not within the range of
applicability of the high-temperature expansion. It is for this
reason that we excluded the region of parameter space yielding a
one-loop zero-temperature potential with a global minimum at the
origin. 

We observe a similar behavior in the region of parameter space where
$M_H\simeq 120$~GeV and $M_\Theta\simeq650$~GeV in the $M_A=
350$~GeV, $M_f= 350$~GeV plot. In this case the black and white
regions cannot be studied via the high-temperature approximation
since $M_\Theta/T_c > 7$, which is a strong
indication of the breakdown of the high-temperature
expansion \footnote{The
difference between black and white regions is due to the fact that in the
black region one still observes a phase transition as function of the
temperature while in the white region no phase transition is found
for any temperature which is a clear indication of the full breakdown
of the high-temperature expansion. Either way these two regions are
not accessible within our approximations.}.
However, we have checked the validity of the high-$T$ expansion for
the other regions of our plots  by adding higher  order terms in the
expansion and seeing how the results change. Including terms up to
and including order $1/T^6$, we find that the quantitative results
presented here are stable against higher order corrections.

\begin{figure}[ht]
 {\includegraphics[height=6.5cm,width=7.82cm]{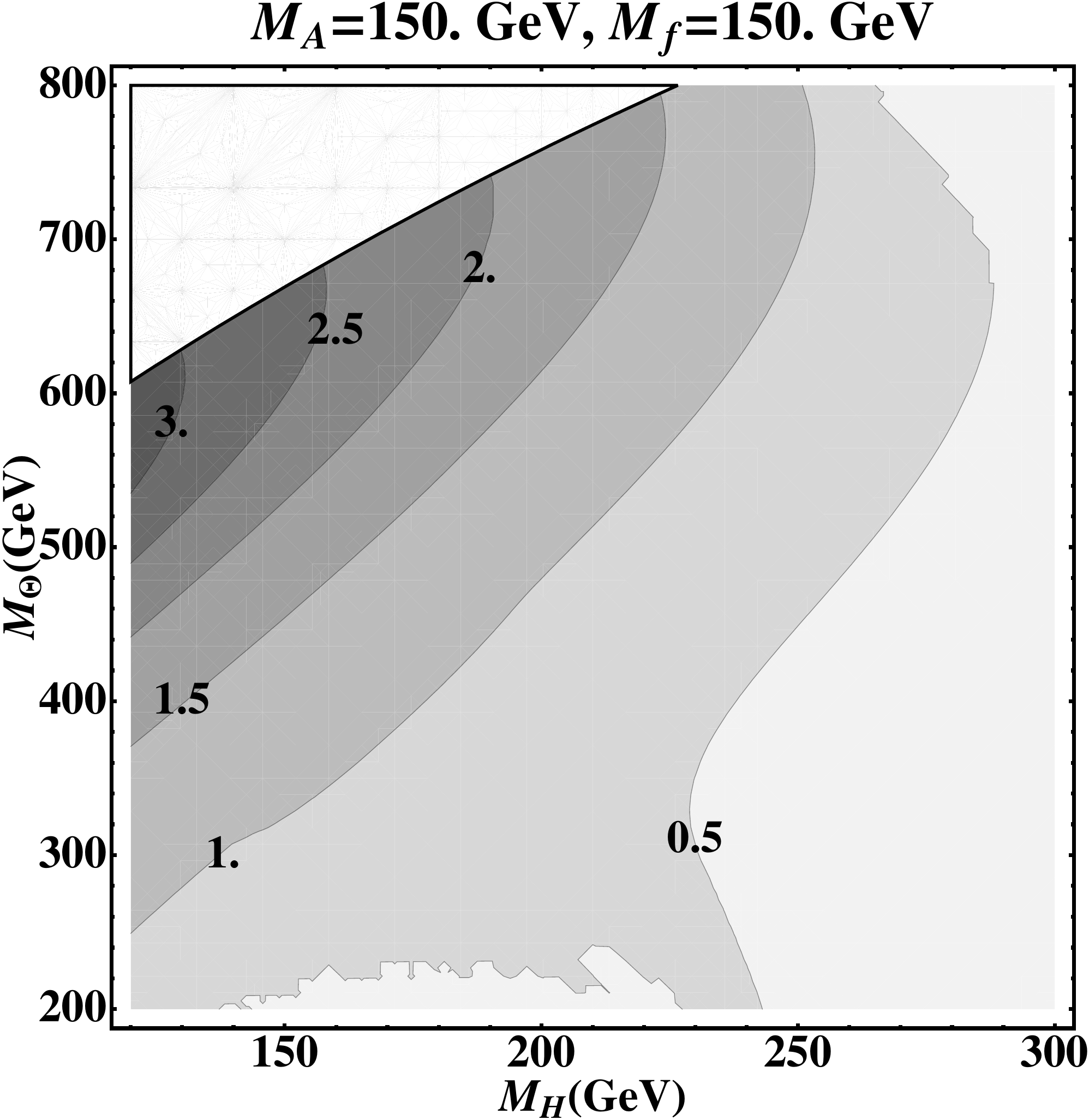}\hspace{0.8cm}\includegraphics[height=6.5cm,width=7.82cm]{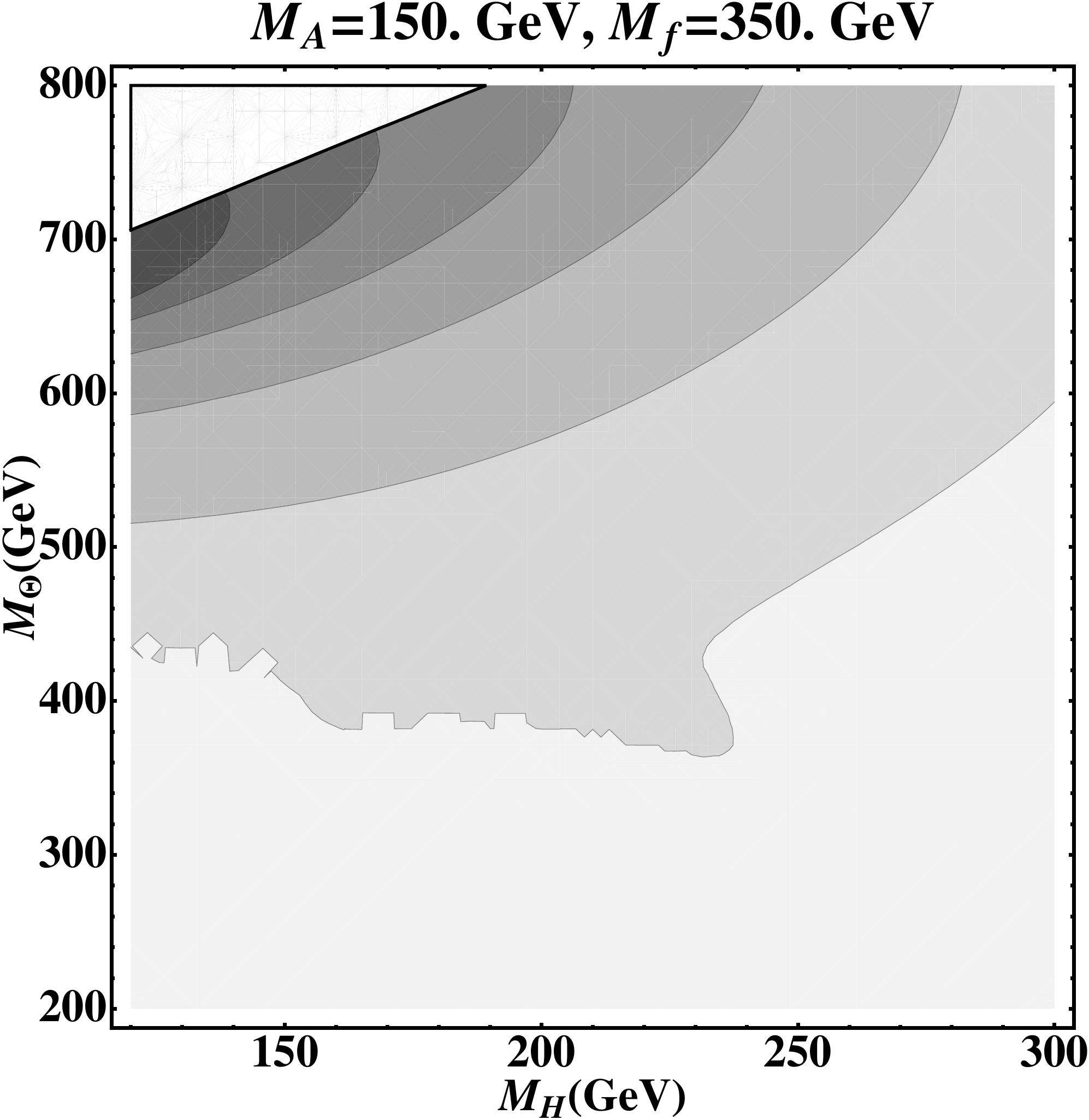}
\includegraphics[height=6.5cm,width=7.82cm]{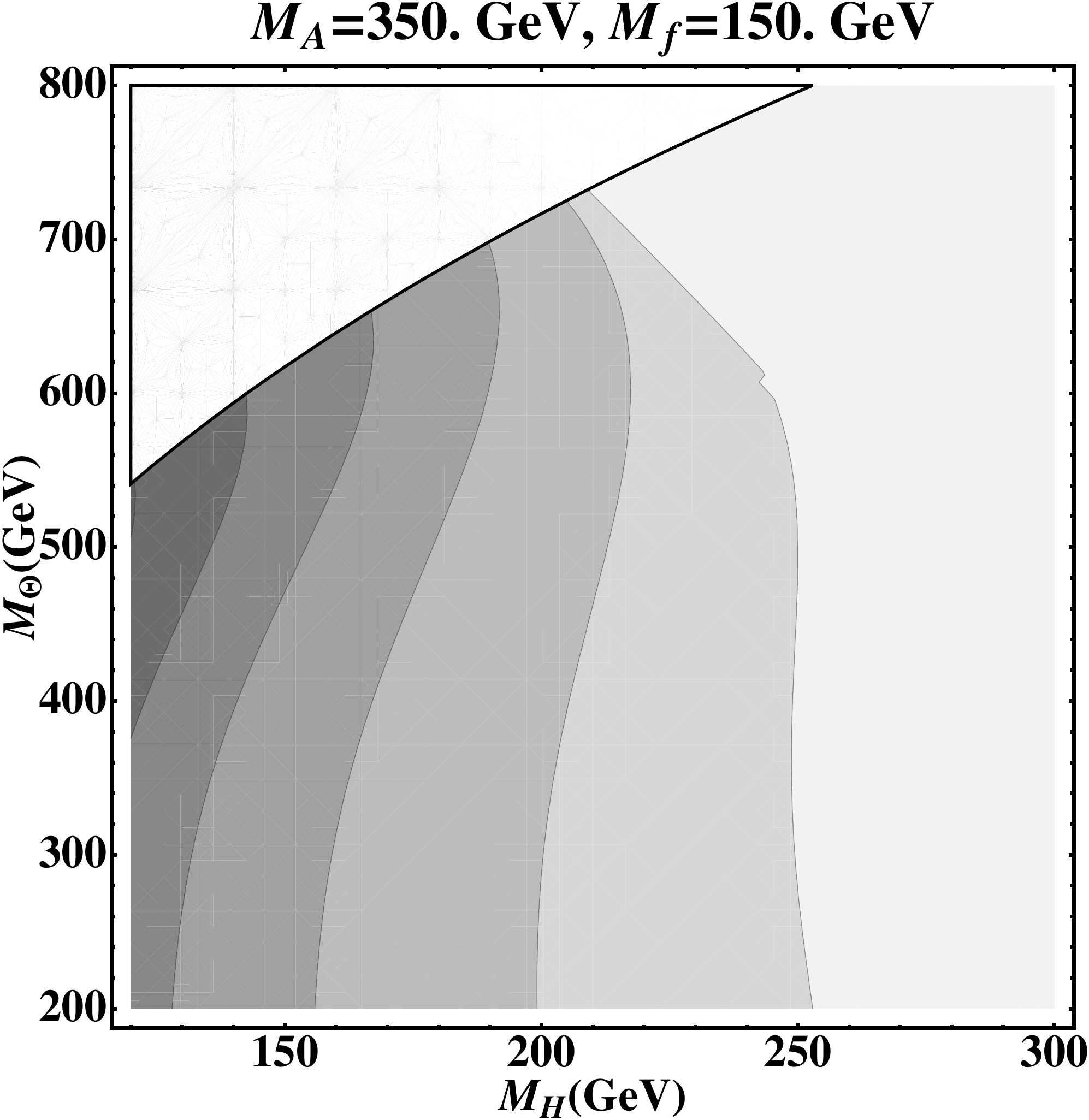}\hspace{0.8cm}\includegraphics[height=6.5cm,width=7.82cm]{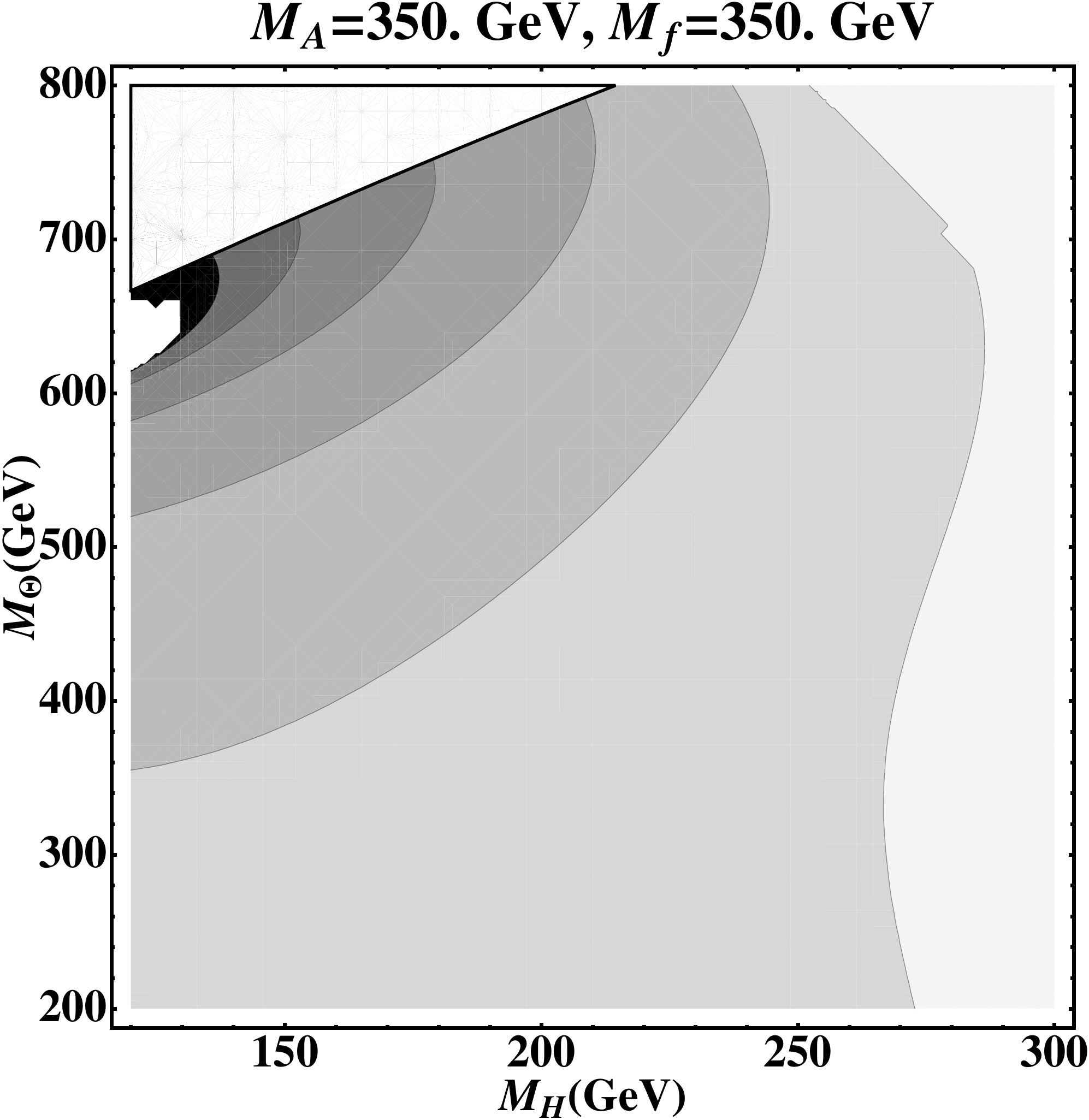}
}
\caption{The strength of the phase transition ($\phi_c/T_c$) in the
$M_H$-$M_{\Theta}$ plane for $M_A$, $M_{\rm f}=150$~GeV and $300$~GeV,
in the heavy ETC mass scenario. $\phi_c/T_c=0.5,1.0,1.5,\ldots3.0$ at
the contour lines, such that $\phi_c/T_c<0.5$ in the region with
lightest color. In the white region in the upper left corners of the
plots the broken phase vacuum is metastable already at
$T=0$.}\label{fig:heavymetc} \end{figure}

\subsection{Light ETC-induced masses}

In the light ETC mass scenario, all of the MWT scalars are relatively
light with respect to the eletroweak scale. Then all the degrees of
freedom which were discussed in Subsection~\ref{zeroTV} are thermally
active at the phase transition.  The strength of the phase transition
in this case is plotted in fig.~\ref{fig:lightmetc}. Since we
found $\phi_c/T_c$ to be 
rather weakly dependent on $m_{\rm ETC}$ when the scalar baryons
are thermally active, we fixed $m_{\rm ETC}=150$~GeV. The transition
is slightly weaker than in the heavy $m_{\rm ETC}$ scenario. For
$M_A\gtrsim 300$~GeV or $M_{\rm f}\gtrsim 500$~GeV no first order
transition is seen. This is why we changed the second reference point
of $M_A$ from $350$~GeV to $250$~GeV in the plots.

\begin{figure}[ht]
 {\includegraphics[height=6.5cm,width=7.82cm]{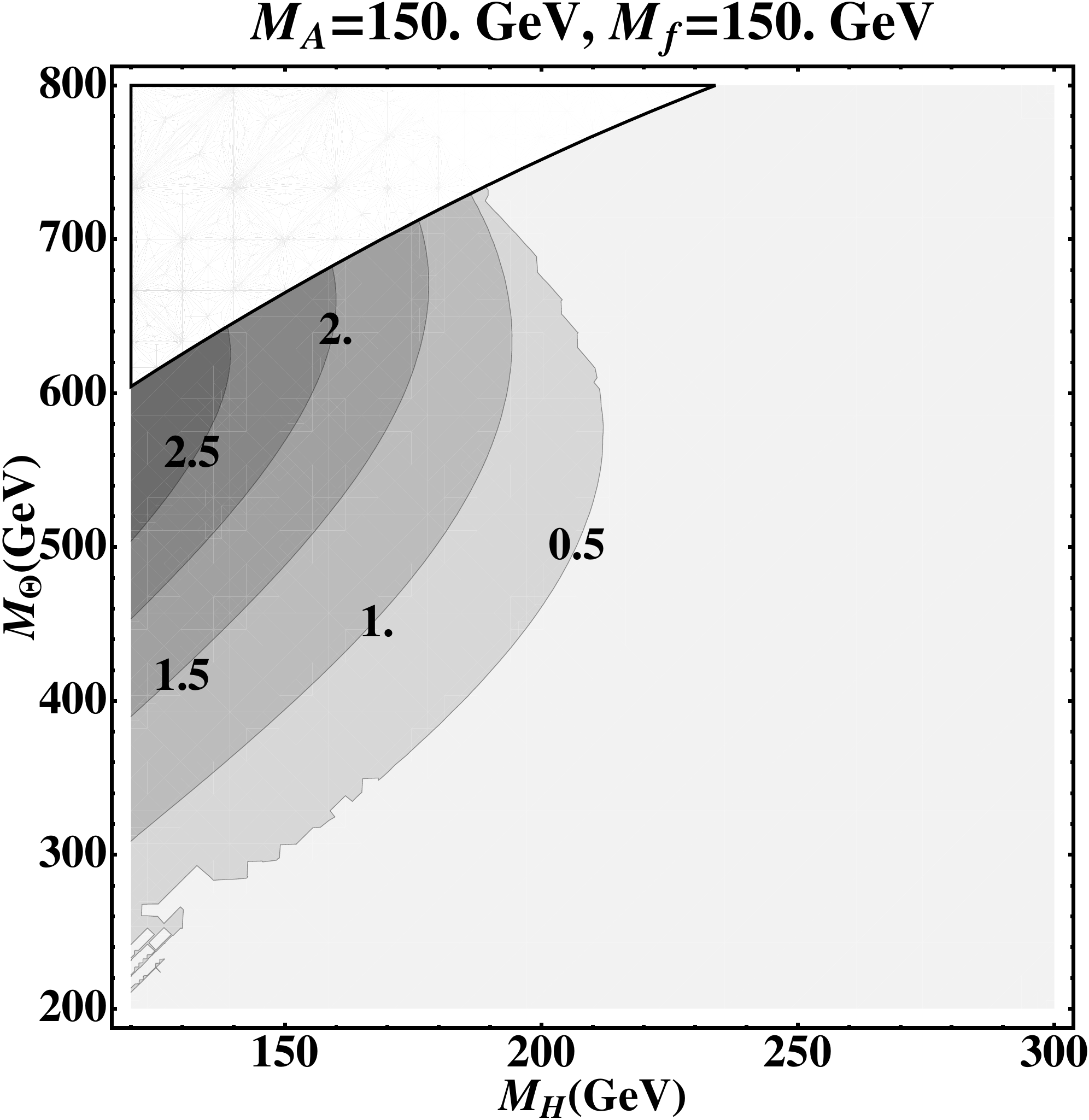}\hspace{0.8cm}\includegraphics[height=6.5cm,width=7.82cm]{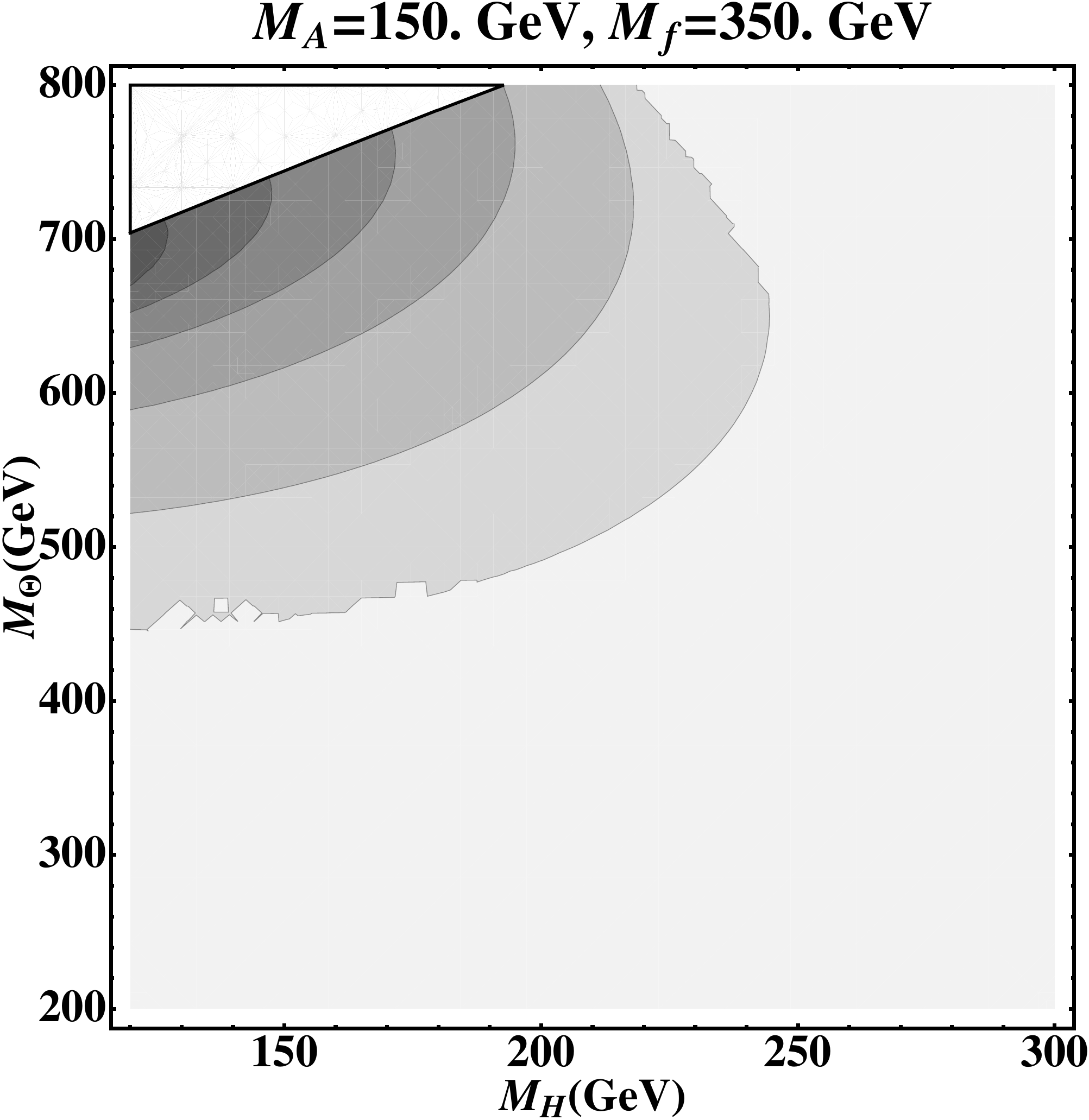}
\includegraphics[height=6.5cm,width=7.82cm]{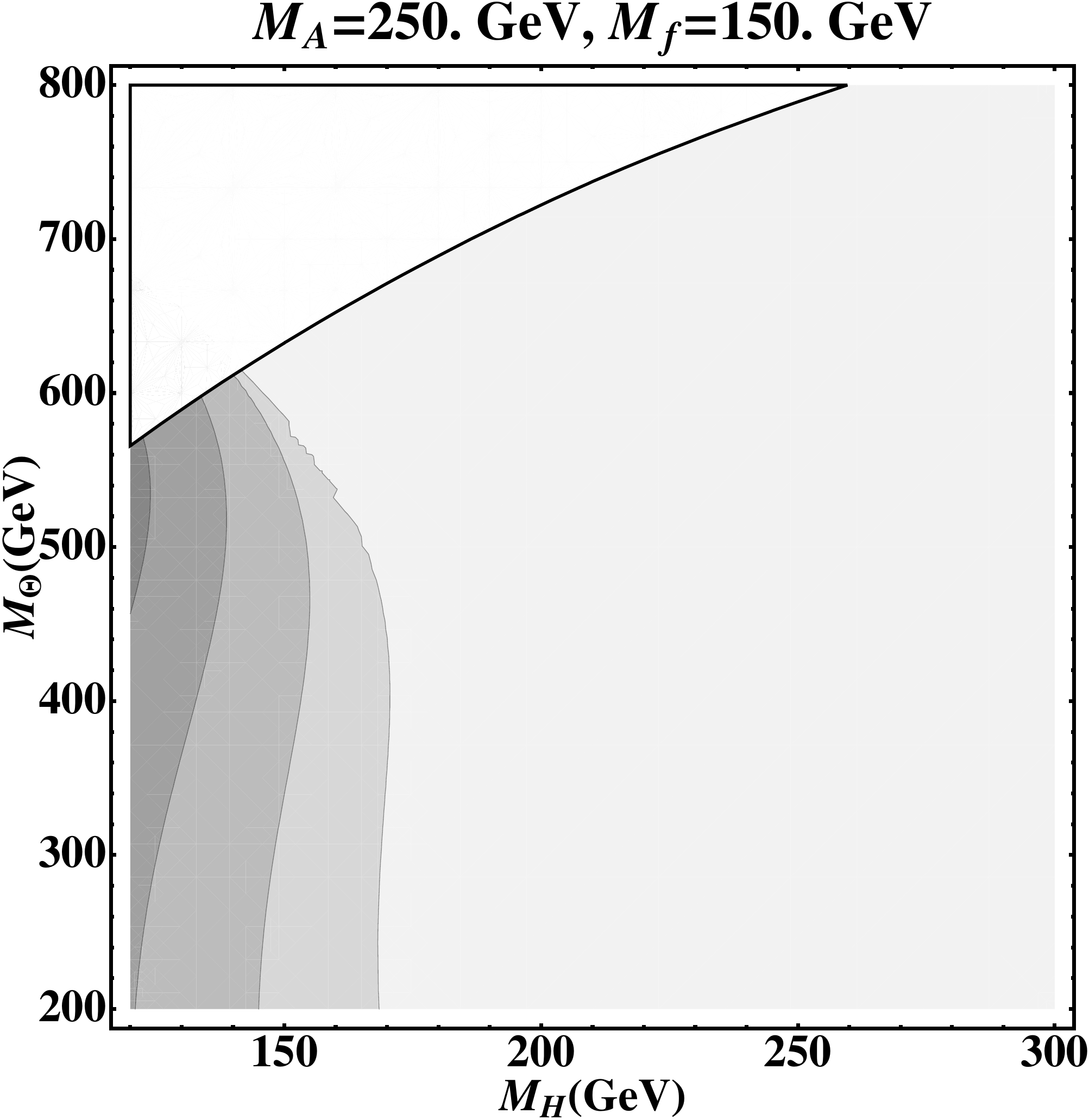}\hspace{0.8cm}\includegraphics[height=6.5cm,width=7.82cm]{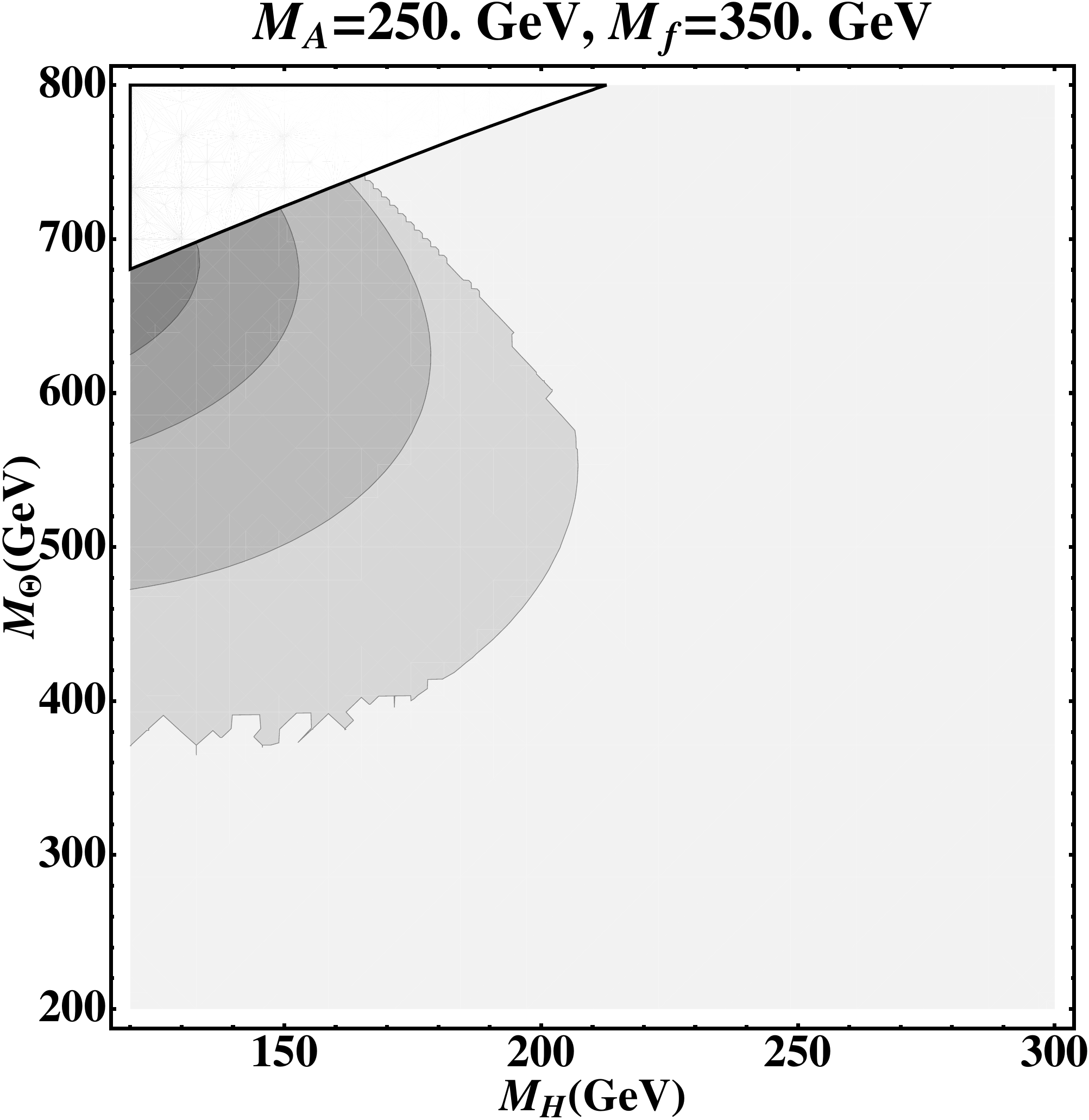}
}
\caption{The strength of the phase transition ($\phi_c/T_c$) in the
$M_H$-$M_{\Theta}$ plane,  for the light ETC mass 
($m_{\rm ETC}=150$~GeV) scenario. $M_A$ and $M_{\rm f}$ are varied 
as indicated in the labels.}\label{fig:lightmetc}
\end{figure}

\subsection{Explanation of Results}

It is possible to qualitatively understand the behavior of
$\phi_c/T_c$ as a function of the masses. In general, a strong first
order phase transition can be achieved if the zero-temperature
potential is close to being flat, {\it i.e.,} the vacuum value
$V(\sigma = v)$  is small and negative (recall that we define
$V(\sigma =0)$ to be zero). Then if the thermal corrections are
strong and positive around $\sigma\simeq v$, the phase transition
takes place at a low temperature, giving a large $\phi_c/T_c$. 

In our model $V(\sigma= v)$ is small typically when the composite 
Higgs mass is low and one of the masses $M_A$ and $M_\Theta$ is a bit
larger. The one-loop zero-temperature contribution of the scalars is
enhanced relative to the tree-level potential for such values of the
masses. It increases the value of the potential at the broken phase
and creates a bump between the two minima.  This is illustrated
in figure \ref{fig3}, which shows the typical shape of the one-loop
correction (\ref{VT0},\ref{fdef}) from bosons, where we have replaced
$\log(m^2(\sigma))$ by $\log(T^2)$ due to the finite-$T$ contribution
canceling this nonanalytic dependence.
Fermions have exactly the
opposite effect,  whence the contributions of the fermions and the
scalar bosons need to be balanced. The baryons do not play a big
role, since their (squared) masses include the hard term $M_{\rm
ETC}^2$ and are thus more weakly dependent on $\sigma$.

\begin{figure}[ht]
 \centerline{\includegraphics[height=6.5cm,width=7.82cm]{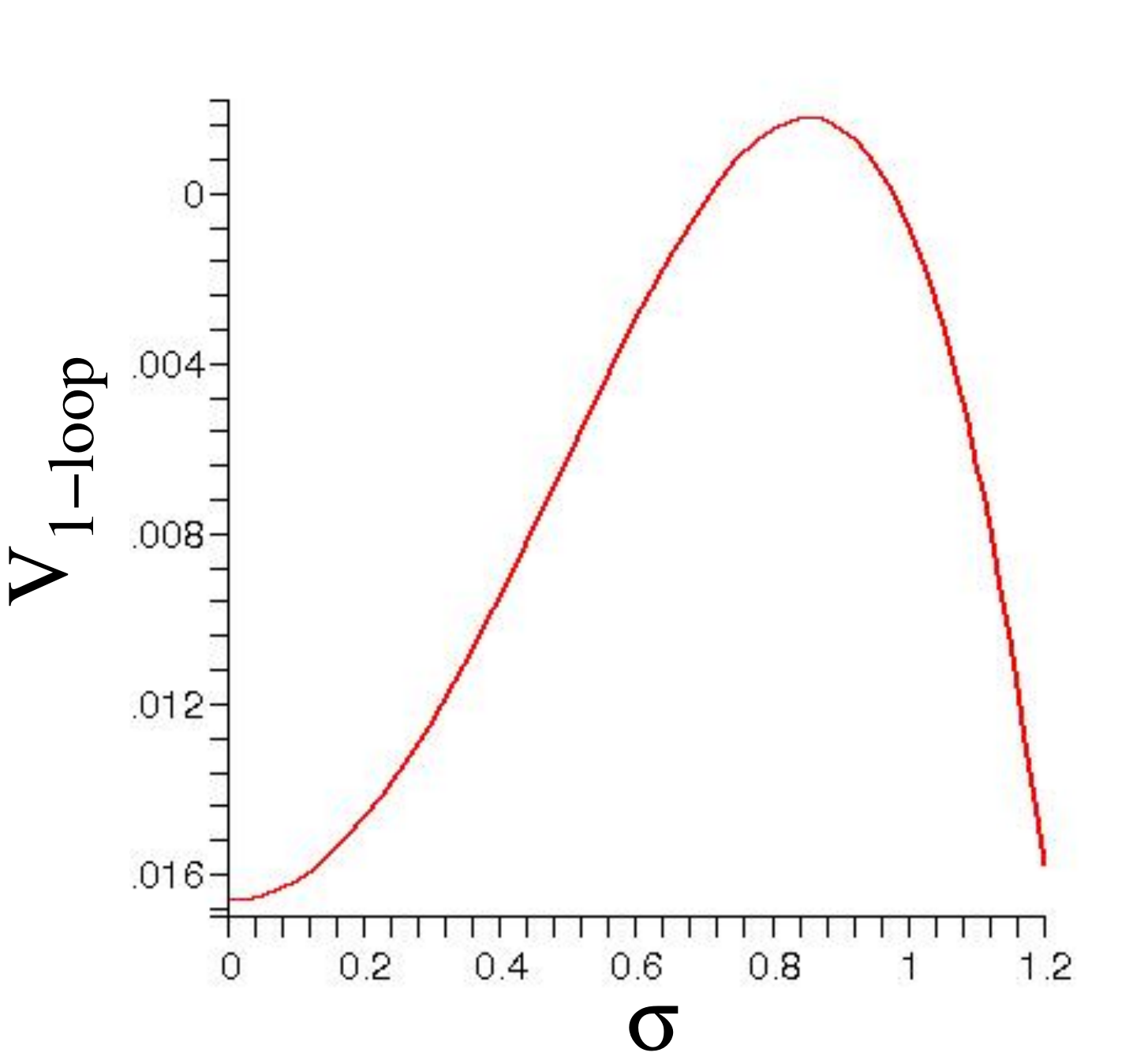}}
\caption{Typical shape of bosonic contribution to one-loop zero
temperature potential, eqs.\ (\ref{VT0},\ref{fdef}), which helps to
enhance the strength of the phase transition.}\label{fig3}
\end{figure}

The shape of the thermal corrections also affects $\phi_c/T_c$, since
a term of the form $-Tm^3\sim -T\phi^3$ creates a barrier between the
symmetric and broken phases in the potential at the critical
temperature, where $V\sim \lambda\phi^2 (\phi - v_c)^2$.  However this is
only true if the field-dependent mass is close to the form $m^2 \sim
g^2\phi^2$.  Thermal and vacuum contributions of the form $m^2 \sim
g^2\phi^2 + m_0^2(T)$ reduce this effect, and for large $m_0^2(T)$,
the expansion of the cubic term in powers of $\phi$ gives
contributions in the wrong direction, tending to reduce $\phi_c/T_c$.

The behavior seen in figs.~\ref{fig:heavymetc}
and~\ref{fig:lightmetc} can be understood as a combination of  the
above effects. In the white regions in the upper left-hand corners of
the plots, the contribution to the bump from the zero-$T$ one-loop
correction is so large that the broken-phase vacuum is metastable
even at $T=0$. Next to this region there is a large part of parameter
space where the scalar and fermion masses are correctly balanced to
produce small and negative  $V(\sigma= v)$, which yields large
$\phi_c/T_c$.   However, in the upper right-hand corner of the plots,
$\sum_b n_b M_b^2$ becomes so large that the thermal corrections do
not enhance the potential at $\sigma \simeq v$ any longer, and the
phase transition is weakened. When the composite baryons are
thermally active (light ETC masses), their effect on the ring
resummation makes the second term of (\ref{VTb}) more negative, but
not sufficiently of the form $\phi^3$. Hence the scalar and Higgs
masses are restricted to be smaller to compensate, which causes the
difference between figs.~\ref{fig:heavymetc} and~\ref{fig:lightmetc}.

Note that when the scalar baryons are decoupled, the scalar fields of
MWT consist of two Higgs doublets. Hence dependences shown in
figs.~\ref{fig:heavymetc} and~\ref{fig:lightmetc} are quite similar
to those of the two Higgs doublet models \cite{Cline:1996mga}. Both
have a metastable broken phase vacuum when the Higgs is light and the
other scalars are heavy with respect to the electroweak scale. The
edge of this region has a similar dependence on $M_H$ in both models.
Strong first order phase transitions are observed near the regions of
metastable vacua in both cases.

{In the parameter region where a strong first order transition is
observed the composite Higgs and its pseudoscalar partner $\Theta$
are light enough to be produced at the LHC. Moreover one expects, for
this range of parameters of the effective Lagrangian, sizable
deviations from the SM predictions at the LHC. {}For example, the
important $pp\rightarrow HW$ process at LHC  is enhanced  relative to
the SM one \cite{Zerwekh:2005wh}. A detailed analysis dedicated to
the LHC phenomenology of nearly-conformal technicolor models is about
to appear \cite{BFFJPS}. We emphasize that the spectrum is completely
fixed by the underlying gauge theory and that first principle lattice
simulations can test our results.}

\subsection{A novel phase transition at lower energies}
\label{sect4.d}
As suggested in \cite{Sannino:2008ha}, an intriguing possibility can
 emerge in that one can have {\em two independent} phase transitions
 at nonzero temperature in technicolor theories, whenever the theory
 possesses a nontrivial center symmetry.  The two phase transitions
 are the chiral one, directly related to the electroweak phase
 transition, and a confining one at lower temperatures.  During the
 history of the universe one predicts a phase transition around the
 electroweak scale and another one at lower temperatures with a jump
 in the entropy proportional to the number of degrees of freedom
 liberated (or gapped) when increasing (decreasing) the temperature
 (see \cite{Mocsy:2003qw} for a simple explanation of this phenomenon
 and a list of relevant references). This may have very interesting
 cosmological consequences. In this work we have concentrated on the
 chiral one alone. The interplay with the confining one, expected to
 occur at lower temperatures, can be studied by coupling the
 effective Lagrangian presented here to the Polyakov-loop effective
 degree of freedom as done in \cite{Mocsy:2003qw}.  %%%%%%%%%%

\acknowledgments
 J.C.\ is supported by
NSERC of Canada. The work of M.J.\ and F.S.\ is supported by the Marie Curie Excellence
Grant under contract MEXT-CT-2004-013510.   We thank the CERN
theory group for its hospitality during the initiation of this work.

\appendix
\section{Zero- and Finite-Temperature Background Dependent Scalar Masses} 
\label{AppA}
The composite scalars are assembled in the matrix $M$ of Eq.~(\ref{M}). In terms of the mass eigenstates this reads
\small
\begin{eqnarray}
M=
\begin{pmatrix}
i \Pi_{UU} + \widetilde{\Pi}_{UU} & {\displaystyle \frac{i \Pi_{UD} + \widetilde{\Pi}_{UD}}{\sqrt{2}}} &
{\displaystyle \frac{\sigma+i\Theta + i \Pi^0 + A^0}{2}} & {\displaystyle \frac{i\Pi^+ + A^+}{\sqrt{2}}} \\
\smallskip \\
{\displaystyle \frac{i \Pi_{UD} + \widetilde{\Pi}_{UD}}{\sqrt{2}}} & i \Pi_{DD} + \widetilde{\Pi}_{DD} &
{\displaystyle \frac{i\Pi^- + A^-}{\sqrt{2}}} & {\displaystyle \frac{\sigma+i\Theta - i \Pi^0 - A^0}{{2}}} \\
\smallskip \\
{\displaystyle \frac{\sigma+i\Theta + i \Pi^0 + A^0}{2}} & {\displaystyle \frac{i\Pi^- + A^-}{\sqrt{2}}} &
i \Pi_{\overline{UU}} + \widetilde{\Pi}_{\overline{UU}} & {\displaystyle \frac{i \Pi_{\overline{UD}} + \widetilde{\Pi}_{\overline{UD}}}{\sqrt{2}}} \\
\smallskip \\
{\displaystyle \frac{i\Pi^+ + A^+}{\sqrt{2}}} & {\displaystyle \frac{\sigma+i\Theta - i \Pi^0 - A^0}{2}} &
{\displaystyle \frac{i \Pi_{\overline{UD}} + \widetilde{\Pi}_{\overline{UD}}}{\sqrt{2}}} & i \Pi_{\overline{DD}} + \widetilde{\Pi}_{\overline{DD}}
\end{pmatrix} \ , \nonumber \\
\end{eqnarray}
\normalsize
where $\sigma=v+H$. The Lagrangian summary for the Higgs sector, including the spontaneously broken potential, and the ETC mass term for the uneaten Goldstone bosons, is
\begin{eqnarray}
{\cal L}_{\rm Higgs} &=& \frac{1}{2}{\rm Tr}\left[D_{\mu}M D^{\mu}M^{\dagger}\right] + \frac{m^2}{2}{\rm Tr}[MM^{\dagger}] \nonumber \\
& - & \frac{\lambda}{4} {\rm Tr}\left[MM^{\dagger} \right]^2 - \lambda^\prime {\rm Tr}\left[M M^{\dagger} M M^{\dagger}\right]
+  2\lambda^{\prime\prime} \left[{\rm Det}(M) + {\rm Det}(M^\dagger)\right] \nonumber \\
& + & \frac{m_{\rm ETC}^2}{4}\ {\rm Tr}\left[M B M^\dagger B + M M^\dagger \right] \ ,
\end{eqnarray}
where the covariant derivative is given by Eq.~(\ref{covariantderivative}).

The zero-temperature background-dependent scalar mass squared eigenstates are: 
\begin{eqnarray} 
-m^2 + m^2_{ETC} + (\lambda - \lambda^{\prime \prime} + \lambda^{\prime}) &\sigma^2& \ , \qquad 6~{\rm degenerate ~states} \nonumber \\ 
-m^2 + m^2_{ETC} + (\lambda +\lambda^{\prime \prime} + 3\lambda^{\prime}) &\sigma^2& \ , \qquad 6~{\rm degenerate ~states} \nonumber \\ 
-m^2 + (\lambda - \lambda^{\prime \prime} + \lambda^{\prime})&\sigma^2&\ , \qquad 3~{\rm degenerate ~states}~ (GB) \nonumber \\
-m^2 + (\lambda + \lambda^{\prime \prime} +3 \lambda^{\prime})&\sigma^2&\ , \qquad 3~{\rm degenerate ~states} \nonumber \\
-m^2 + 3(\lambda - \lambda^{\prime \prime} + \lambda^{\prime})&\sigma^2&\ , \qquad 1~{\rm state~ (Higgs)} \nonumber \\
-m^2 + (\lambda + 3 \lambda^{\prime \prime} + \lambda^{\prime})&\sigma^2&\ , \qquad 1~{\rm state} 
\end{eqnarray}
The temperature-dependent (one-loop) effective scalar masses of the 
Arnold-Espinosa approximation \cite{Arnold:1992rz} are calculated as 
follows. Compute the $T^2$ term of the one-loop thermal 
correction $V_T^{(1)}$ as explained in subsection~\ref{sec:oneloopT}, but in an arbitrary background, {\it i.e.}, function of  all the scalar fields. 
Then, for example, the contribution of the top quark loop reads
\begin{eqnarray}
 {V_{T^2}^{(1)}}_{\rm Top} = \frac{T^2}{4}\left.M_{\rm Top}^2\right|_{\rm background} = \frac{m_{\rm Top}^2 T^2
   \left(\left(\Theta+\Pi^0\right){}^2+\left(\sigma+A^0\right){}^2\right)}{4 v^2} \ .
\end{eqnarray}
The effective thermal masses are obtained by adding to the 
$T=0$ scalar mass matrix the thermal mass matrix
\begin{eqnarray}
 M_{ij} = \frac{\partial^2}{\partial v_i \partial v_j} V_{T^2}^{(1)} \ .
\end{eqnarray}
Here $v_i$ represents the i-th scalar field thermally active at the electroweak phase transition. 

The one-loop finite-temperature correction to the scalar masses, due
solely to the scalar self-interactions, and considering all of
the 20 bosons to be thermally active, is
\begin{equation}
\frac{T^2}{6}(11\lambda + 20 \lambda^{\prime}) \ .
\end{equation}
However the full finite-temperature corrections have involved expressions when taking into account all the particles, {\it i.e.,} gauge bosons and fermions. 

We summarize below the temperature and background dependent scalar masses in the case in which the ETC states are heavy,  and hence integrated out, and the case in which we retain all of the states. 

\subsection{Heavy $M_{etc}$}
\begin{eqnarray}
M_{\Pi^\pm}^2(\sigma,T)&=&-m^2+(\lambda  +\lambda '-\lambda '' ) \sigma ^2+\frac{T^2}{6}  (5 \lambda +8 \lambda ' )+\frac{T^2}{16} \left(g_1^2+3 g_2^2\right)\\
M_{A^\pm}^2(\sigma,T)&=&-m^2+(\lambda  +3 \lambda ' +\lambda '') \sigma ^2+\frac{T^2}{6}  (5 \lambda +8 \lambda ' )+\frac{T^2}{16} \left(g_1^2+3 g_2^2\right)\end{eqnarray}
\begin{eqnarray}
M_{\Theta/\Pi^0}^2(\sigma,T)&=&
-m^2+(\lambda +\lambda '+\lambda ''  ) \sigma ^2\nonumber\\&&+\frac{T^2}{6} (5 \lambda +8 \lambda ' )+\frac{T^2}{16} \left(g_1^2+3 g_2^2\right)+\frac{T^2}{6 v^2} \left(m_E^2+m_N^2+3 m_{\text{Top}}^2\right)\nonumber\\&&\pm\sqrt{\left(2 \lambda '' \sigma ^2\right)^2+\left[\frac{T^2}{6 v^2}\left(-m_E^2+m_N^2+3 m_{\text{Top}}^2\right)\right]^2 }\\
M_{H/A^0}^2(\sigma,T)&=&-m^2+(2 \lambda +3 \lambda '-\lambda ''  ) \sigma ^2\nonumber\\&&+\frac{T^2}{6} (5 \lambda +8 \lambda ' )+\frac{T^2}{16} \left(g_1^2+3 g_2^2\right)+\frac{T^2}{6 v^2} \left(m_E^2+m_N^2+3 m_{\text{Top}}^2\right)\nonumber\\&&\pm\sqrt{\left[(\lambda -2 \lambda '' )\sigma ^2\right]^2+\left[\frac{T^2}{6 v^2}\left(-m_E^2+m_N^2+3 m_{\text{Top}}^2\right)\right]^2 } 
\end{eqnarray}

\subsection{
Light $M_{etc}$}
\begin{eqnarray}
M_{\Pi^\pm}^2(\sigma,T)&=&-m^2+(\lambda  +\lambda '-\lambda '' ) \sigma ^2+\frac{T^2}{6}   (11 \lambda +20 \lambda ' ) +\frac{T^2}{16} \left(g_1^2+3 g_2^2\right)\\
M_{A^\pm}^2(\sigma,T)&=&-m^2+(\lambda  +3 \lambda ' +\lambda '') \sigma ^2+\frac{T^2}{6}   (11 \lambda +20 \lambda ' ) +\frac{T^2}{16} \left(g_1^2+3 g_2^2\right)\end{eqnarray}
\begin{eqnarray}
M_{\Theta/\Pi^0}^2(\sigma,T)&=&
-m^2+(\lambda +\lambda '+\lambda ''  ) \sigma ^2\nonumber\\&&+\frac{T^2}{6}  (11 \lambda +20 \lambda ' ) +\frac{T^2}{16} \left(g_1^2+3 g_2^2\right)+\frac{T^2}{6 v^2} \left(m_E^2+m_N^2+3 m_{\text{Top}}^2\right)\nonumber\\&&\pm\sqrt{\left(2 \lambda '' \sigma ^2\right)^2+\left[\frac{T^2}{6 v^2}\left(-m_E^2+m_N^2+3 m_{\text{Top}}^2\right)\right]^2 }\\
M_{H/A^0}^2(\sigma,T)&=&-m^2+(2 \lambda +3 \lambda '-\lambda ''  ) \sigma ^2\nonumber\\&&+\frac{T^2}{6}  (11 \lambda +20 \lambda ' ) +\frac{T^2}{16} \left(g_1^2+3 g_2^2\right)+\frac{T^2}{6 v^2} \left(m_E^2+m_N^2+3 m_{\text{Top}}^2\right)\nonumber\\&&\pm\sqrt{\left[(\lambda -2 \lambda '' )\sigma ^2\right]^2+\left[\frac{T^2}{6 v^2}\left(-m_E^2+m_N^2+3 m_{\text{Top}}^2\right)\right]^2 }   \end{eqnarray}

\begin{eqnarray}
M_{\Pi_{UD}/\widetilde{\Pi}_{UD}}^2(\sigma,T)&=&-m^2+m_{\rm ETC}^2+ (\lambda  +2\lambda ' ) \sigma ^2+\frac{T^2}{6}   (11 \lambda +20 \lambda ' ) +\frac{T^2}{4} \left(y^2 g_1^2+ g_2^2\right)\nonumber\\&& \pm \sqrt{\left[(\lambda ' + \lambda '' )\sigma ^2\right]^2+\left(\frac{1}{4}T^2g_2^2\right)^2 } \\
M_{\Pi_{UU}/\widetilde{\Pi}_{UU}}^2(\sigma,T)&=&-m^2+m_{\rm ETC}^2+ (\lambda  +2\lambda ' ) \sigma ^2+\frac{T^2}{6}   (11 \lambda +20 \lambda ' ) \nonumber \\ &&+\frac{T^2}{8} \left[\left(1+2y +2y^2\right) g_1^2+ 2 g_2^2\right]
\nonumber\\&& 
\pm \sqrt{\left[(\lambda ' + \lambda '' )\sigma ^2\right]^2+\left\{\frac{T^2}{8} \left[\left(1+2y\right) g_1^2- 2 g_2^2\right]\right\}^2}  \\
M_{\Pi_{DD}/\widetilde{\Pi}_{DD}}^2(\sigma,T)&=&-m^2+m_{\rm ETC}^2+ (\lambda  +2\lambda ' ) \sigma ^2 +\frac{T^2}{6}   (11 \lambda +20 \lambda ' ) \nonumber \\&&+\frac{T^2}{8} \left[\left(1-2y +2y^2\right) g_1^2+ 2 g_2^2\right]\nonumber\\&& \pm \sqrt{\left[(\lambda ' + \lambda '' )\sigma ^2\right]^2+\left\{\frac{T^2}{8} \left[\left(1-2y\right) g_1^2- 2 g_2^2\right]\right\}^2}  
\end{eqnarray} 
Here the notation  $M_{A/B}$ means that the states $A$ and $B$ are mixed through thermal corrections.  The diagonal thermal masses of each $A/B$ system are reported on the right hand side.

\section{Transverse and Longitudinal Gauge Boson Mass Matrix}
\label{AppB}

The background dependent transverse gauge boson mass matrix is:
\begin{equation}
M_T^2(\sigma) = \frac{\sigma^2}{4}
\left[
\begin{array}{cccc}
g^2  & 0  & 0 & 0  \\
0  & g^2  &0    &0 \\
0  & 0   & g^2  & -g^{\prime}g\\
 0 & 0& -g^{\prime}g & {g^{\prime}}^2
\end{array}
\right] \ ,
\end{equation}
while the longitudinal background dependent Debye mass is in MWT:
\begin{equation}
M_L^2(\sigma) = M_T^2(\sigma)+\Pi_L \ ,\end{equation} with \begin{equation}\Pi_L =\left[
\begin{array}{cccc}
(2+\frac{5}{6}) g^2 T^2& 0  & 0 & 0  \\
0  & (2+\frac{5}{6}) g^2 T^2  &0    &0 \\
0  & 0   & (2+\frac{5}{6}) g^2 T^2&0\\
 0 & 0& 0 & f(y){g^{\prime}}^2 T^2
\end{array}
\right] \ ,
\end{equation}
and $3f(y) = 1+(6y^2+2)+5+\frac{1}{2}(9y^2+\frac{1}{2}) + \frac{1}{2}(y^2 + \frac{1}{2})$. The longitudinal mass matrix receives finite temperature contributions from the scalars, the new lepton family, the techniquark-technigluon states which adds to the usual SM corrections  (but with the standard Higgs replaced by the technicolor sector).  
The transverse bosons acquire a magnetic mass of order $g^2T$ which we have neglected. To compute the nonzero temperature corrections to the longitudinal vector boson masses we have used the formulae: 
\begin{eqnarray}
\underline{U(1)} \qquad \qquad \Pi_L ^{S} = \frac{{g^{\prime}}^2T^2}{3} \sum_{S} Y^2_{S} \ , \qquad \qquad  \Pi_L ^{F} = \frac{{g^{\prime}}^2T^2}{6} \sum_{F} Y^2_{F} 
\end{eqnarray}
where the sums are over complex scalars and chiral fermions respectively. For the nonabelian part we used:
\begin{eqnarray}
\underline{SU(N)} \qquad \Pi_L ^{S} = \frac{{g}^2T^2}{3} \sum_{S} t_2(R_S) \ ,  \qquad  \Pi_L ^{F} = \frac{{g}^2T^2}{6} \sum_{F}  t_2(R_F) \ ,
\qquad \Pi_L ^{V} = \frac{N}{3}{g}^2T^2 \ ,\end{eqnarray}
where $\delta^{ab} t_2(R) = {\rm Tr}\left[ T^a T^b\right]$. 
It is instructive to separate the various contributions to $\Pi_L$. 

{}For the $U(1)$ part the fermionic and scalar contributions read:
\begin{eqnarray}
\Pi_L^{F}& =& T^2  {g^{\prime}}^2 \left[ \frac{5}{9}N_g + \frac{18y^2 + 1}{12}  + \frac{2y^2 + 1}{12}\right]    \ , \\
\Pi_L^{S}& = & \frac{T^2  {g^{\prime}}^2}{3} \left[1 +  (6y^2 + 2)\right] 
\end{eqnarray}
where the first contribution counts $N_g = 3$ generations of the SM fermions, the second contribution is due to the new non technicolor family while the last term is due to the techniquark-technigluon fermion states. For the bosonic sector the first term is due to the two Higgs doublets contained in ${\cal O}$ while the term in brackets takes into account the contribution of the other di-techniquark type of states.

{}For the SU(2) part the fermionic, scalar and vector contributions reads:
\begin{eqnarray}
\Pi_L^{F}& =& \frac{T^2  {g}^2}{6} \left[ 2 N_g + \frac{1}{2}  + \frac{1}{2}\right]    \ , \\
\Pi_L^{S}& = & \frac{T^2  {g^2}}{3} \left[1 +  2\right] \ ,\\
\Pi_L^{V}& = & \frac{2}{3}T^2  {g^2} \ . 
\end{eqnarray}
In the fermionic case the first contribution counts $N_g = 3$ generations of the SM fermions, the second contribution is due to the new non technicolor family while the last term is due to the techniquark-technigluon fermion states. For the bosonic sector the first term is due to the two Higgs doublets contained in ${\cal O}$ while the term in brackets takes into account the contribution of the other di-techniquark type of states (the ones in the upper left component of the matrix $M$).

\section{Generators\label{appgen}}

It is convenient to use the following representation of SU(4)
\beq S^a = \begin{pmatrix} \bf A & \bf B \\ {\bf B}^\dag & -{\bf A}^T
\end{pmatrix} \ , \qquad X^i = \begin{pmatrix} \bf C & \bf D \\ {\bf
    D}^\dag & {\bf C}^T \end{pmatrix} \ , \eeq
where $A$ is hermitian, $C$ is hermitian and traceless, $B = -B^T$ and
$D = D^T$. The ${S}$ are also a representation of the $SO(4)$
generators, and thus leave the vacuum invariant $S^aE + E {S^a}^T = 0\ $.
Explicitly, the generators read
\beq S^a = \frac{1}{2\sqrt{2}}\begin{pmatrix} \tau^a & \bf 0 \\ \bf 0 &
  -\tau^{aT} \end{pmatrix} \ , \quad a = 1,\ldots,4 \ , \eeq
where $a = 1,2,3$ are the Pauli matrices and $\tau^4 =
\mathbbm{1}$. These are the generators of SU$_V$(2)$\times$ U$_V$(1).
\beq S^a = \frac{1}{2\sqrt{2}}\begin{pmatrix} \bf 0 & {\bf B}^a \\
{\bf B}^{a\dag} & \bf 0 \end{pmatrix} \ , \quad a = 5,6 \ , \eeq
with
\beq B^5 = \tau^2 \ , \quad B^6 = i\tau^2 \ . \eeq
The rest of the generators which do not leave the vacuum invariant are
\beq X^i = \frac{1}{2\sqrt{2}}\begin{pmatrix} \tau^i & \bf 0 \\
\bf 0 & \tau^{iT} \end{pmatrix} \ , \quad i = 1,2,3 \ , \eeq
and
\beq X^i = \frac{1}{2\sqrt{2}}\begin{pmatrix} \bf 0 & {\bf D}^i \\
{\bf D}^{i\dag} & \bf 0 \end{pmatrix} \ , \quad i = 4,\ldots,9 \ ,
\eeq
with
\beq\begin{array}{r@{\;}c@{\;}lr@{\;}c@{\;}lr@{\;}c@{\;}l}
D^4 &=& \mathbbm{1} \ , & \quad D^6 &=& \tau^3 \ , & \quad D^8 &=& \tau^1 \ , \\
D^5 &=& i\mathbbm{1} \ , & \quad D^7 &=& i\tau^3 \ , & \quad D^9 &=& i\tau^1
\ .
\end{array}\eeq

The generators are normalized as follows
\beq {\rm Tr}\left[S^aS^b\right] =\frac{1}{2}\delta^{ab}\ , \qquad \ , {\rm Tr}\left[X^iX^j\right] =
\frac{1}{2}\delta^{ij} \ , \qquad {\rm Tr}\left[X^iS^a\right] = 0 \ . \eeq

\end{document}